\documentclass{article}
\usepackage[preprint]{neurips_2026}

\usepackage[utf8]{inputenc} 
\usepackage[T1]{fontenc}    
\usepackage{hyperref}       
\usepackage{url}            
\usepackage{booktabs}       
\usepackage{amsfonts}       
\usepackage{nicefrac}       
\usepackage{microtype}      
\usepackage{xcolor}         

\usepackage{courier}
\usepackage{graphicx}
\usepackage{algorithm}
\usepackage{algorithmic}
\usepackage{fontspec}
\usepackage{makecell}
\usepackage{braket}
\usepackage{booktabs}
\usepackage{fancyhdr}
\usepackage{amsmath, amssymb, bm}
\usepackage{wrapfig}
\usepackage{kotex}
\usepackage{multirow}
\usepackage{booktabs}
\usepackage{amsthm} 

\newtheorem{proposition}{Proposition}[section]

\def\vtheta{{\bm{\theta}}}
\def\va{{\bm{a}}}

\title{Neural Quantum Spectral Operator Learning for Solving Partial Differential Equations}

\author{%
  Chanyoung Kim\textsuperscript{1} \And
  Myeonghwan Seong\textsuperscript{2} \And
  Yujin Kim\textsuperscript{2} 
  \AND
  Daniel K. Park\textsuperscript{2,3,4}\!\,\setcounter{footnote}{1}\thanks{Corresponding authors: Daniel K. Park (dkd.park@yonsei.ac.kr), Youngjoon Hong (hongyj@snu.ac.kr)} \And
  Youngjoon Hong\textsuperscript{5,$\dagger$} 
  \AND
  \\[-0.5cm]
  \normalfont 
  \begin{tabular}{c}
    \textsuperscript{1}Department of Mathematical Sciences, KAIST \\
    \textsuperscript{2}Department of Statistics and Data Science, Yonsei University \\
    \textsuperscript{3}Department of Applied Statistics, Yonsei University \\
    \textsuperscript{4}Department of Quantum Information, Yonsei University \\
    \textsuperscript{5}Department of Mathematical Sciences, Seoul National University \\
    [0.2cm]
    \texttt{chan00young@kaist.ac.kr} \\
    \texttt{\{europa0306, k.yujin2228, dkd.park\}@yonsei.ac.kr} \\
    \texttt{hongyj@snu.ac.kr}
  \\[-0.4cm]
  \end{tabular}
}

\begin{document}

\maketitle

\begin{abstract}
Partial differential equations (PDEs) are central to modeling physical and engineering systems, but repeatedly solving parametric PDEs remains computationally expensive. Operator learning enables fast surrogate inference, yet typically requires large input–output paired datasets generated by costly high-fidelity PDE solvers. Unsupervised operator learning frameworks alleviate data dependency but remain hindered by computational bottlenecks. To address this, we propose Neural Variational Quantum Linear Solver (NVQLS), the first hybrid quantum–classical operator learning framework leveraging the Legendre--Galerkin weak formulation. We critically resolve the sign ambiguity in VQLS energy minimization, preventing erroneous solution representations. Additionally, we introduce a neural embedding, a novel encoding scheme to map varying forcings and PDE coefficients into parameterized quantum circuit representations. These structural innovations provide theoretical computational complexity advantages under efficient state preparation schemes, while achieving superior accuracy compared to a representative classical baseline. Validations on 1D and 2D parametric PDEs under diverse boundary conditions demonstrate NVQLS's capability to simultaneously process varying inputs, offering a scalable unsupervised approach to quantum-enhanced operator learning.
\end{abstract}

\section{Introduction}

    Solving partial differential equations (PDEs) lies at the heart of scientific computing, underpinning advances in physics~\cite{doi:10.1126/sciadv.1602614, 10.1063_5.0188830} and engineering~\cite{make7040137, doi:10.1126/sciadv.abk0644, Koric_s00366-023-01822-x} as well as downstream tasks such as inverse problems~\cite{LEE2026113665, cho2025physicsinformeddeepinverseoperator} and uncertainty quantification~\cite{Zou22M1518189, guo2023ibuqinformationbottleneckbased}. Operator learning~\cite{lu2021learning, 10.5555_3648699.3648788, 10.5555_3540261.3542102} aims to approximate the PDE operator, which maps PDE inputs (e.g., forcing, coefficients, boundary conditions) to corresponding PDE solutions. 
    Models such as FNO~\cite{li2020} and DeepONet~\cite{lu2021learning} enable fast inference once trained, but typically require large collections of precomputed solutions. By incorporating PDE information directly into the objective function, unsupervised operator learning~\cite{PIDON, li2021, cmame_choi2024} bypasses this reliance but still remains computationally bottlenecked as the system size grows. For instance, sampling-based frameworks utilizing automatic differentiation (AD), such as PI-DON~\cite{PIDON}, suffer from diminished sampling efficiency and slower convergence as the spatial dimension $d$ increases~\cite{pmlr-v242-wang24b, MENON2025118403, HU2024106369, 10.1007/s10915-024-02700-4, kharazmi2019variationalphysicsinformedneuralnetworks}.
    Meanwhile, although weak-formulation approaches operating in coefficient space (e.g., SCLON~\cite{cmame_choi2024}) provide high accuracy by leveraging enriched numerical schemes, they struggle with exponentially expanding system sizes in high-dimensional settings with a large basis expansion size $N$.

    Quantum computing offers a distinct computational paradigm in scientific computing, providing a potential route to mitigating classical computational bottlenecks. For example, the Harrow--Hassidim--Lloyd (HHL) algorithm~\cite{HHL} provides a theoretical framework for solving linear systems with runtime polynomial in $\log N$ under strong assumptions. 
    Variational Quantum Linear Solvers (VQLS)~\cite{bravo2019variational} were later proposed as heuristic alternatives that avoid costly subroutines 
    by instead employing parameterized quantum circuits and classical optimization. 
    With standard PDE discretization schemes, these quantum linear solvers have emerged as powerful tools for addressing various PDE problems \cite{morales2025, ali2023performance, liu2021variational, e25040580, Quantum_Spectral_Methods_for_Differential_Equations}. 
    Parallel to these numerical solvers, learning-based paradigms \cite{PhysRevA.110.022612, Kyriienko2021}, including Quantum PINNs \cite{e26080649, Sedykh_2024} and QFNO \cite{jain2024quantum} leverage variational quantum circuits as generic function approximators for modeling PDE solutions. However, VQLS and PINN-type solvers are inherently instance-specific, requiring re-optimization for every new PDE condition. While existing quantum operator frameworks, such as Quantum DeepONet \cite{xiao2025quantum}, can generalize across varying inputs, they predominantly depend on supervised, data-driven training that requires massive datasets of precomputed solutions. These constraints limit the scalability of quantum PDE solvers in high-dimensional and high-resolution scenarios.

    To overcome these limitations, we propose Neural Variational Quantum Linear Solver (NVQLS), a hybrid quantum–classical framework for unsupervised operator learning. To the best of our knowledge, this is the first quantum spectral operator learning framework to incorporate the Legendre-Galerkin weak formulation into its loss construction. Our main contributions are summarized as follows:

\begin{itemize}
    \item We introduce NVQLS, which integrates the Legendre--Galerkin weak formulation with a VQLS-inspired objective to learn solution operators across PDE instances without requiring precomputed solution labels.
    
    \item We develop a phase-aware overlap objective that resolves the sign ambiguity of standard VQLS. This is critical for PDE applications, where sign-flipped quantum state representatives can lead to physically and numerically incorrect classical solution fields.

    \item We introduce a neural embedding scheme that maps varying PDE inputs, including forcing functions and coefficients, to quantum circuit parameters. This enables multi-instance operator learning with shallow variational circuits, instead of retraining a separate quantum model for each PDE instance.

    \item We provide a complexity analysis showing improved per-iteration time and classical memory scaling over classical unsupervised neural-operator baselines, and validate NVQLS on diverse 1D and 2D PDEs, including jointly varying forcing--coefficient inputs and a wave-equation benchmark where NVQLS outperforms a representative unsupervised classical baseline.

\end{itemize}

\begin{figure}[t]
\begin{center}
    \includegraphics[width=\linewidth]{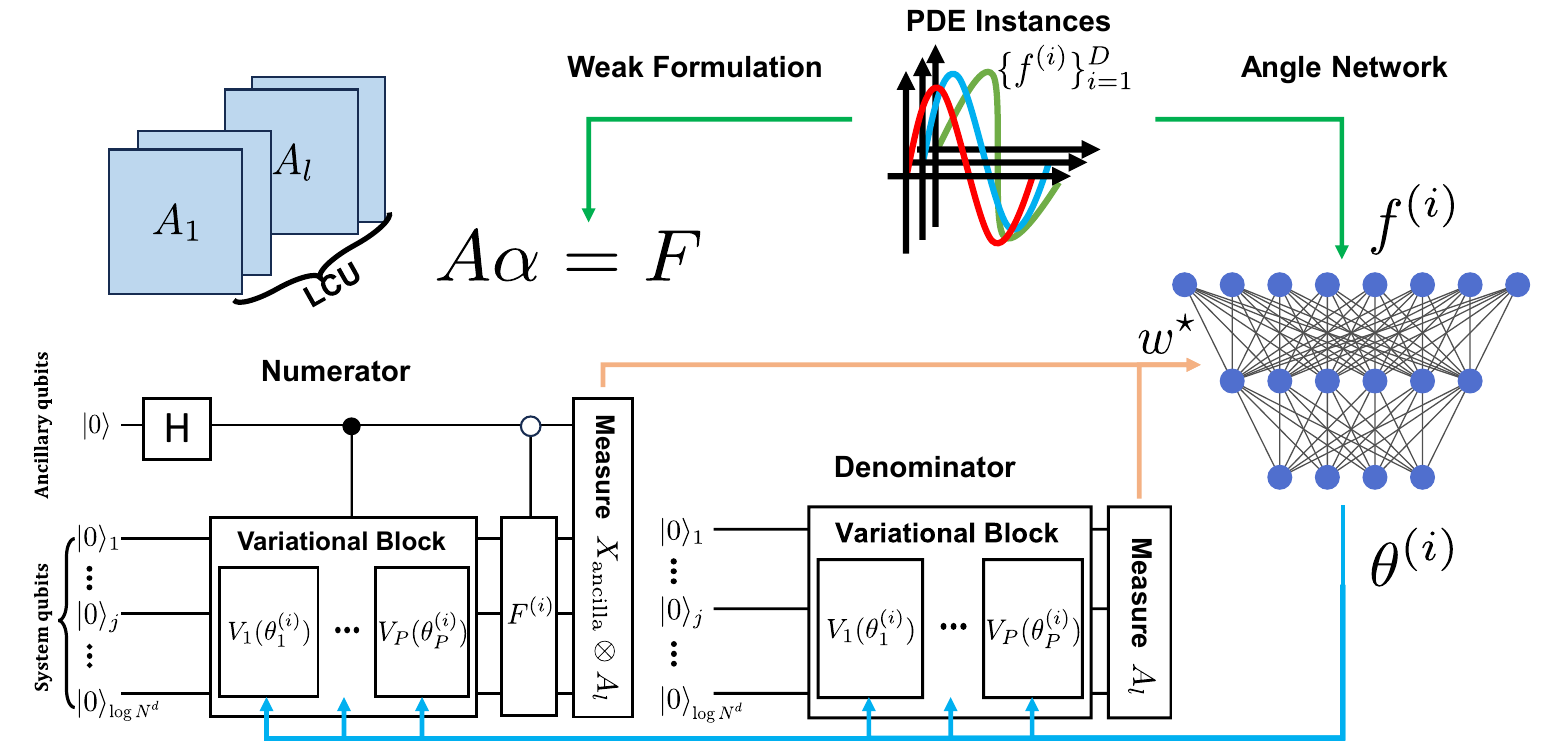}
\end{center}
    \caption{
    Training procedure of the NVQLS framework. An angle network encodes a batch of PDE instances into quantum circuit parameters $\theta$, while a loss function based on quantum state overlap is utilized to minimize weak-form residuals. Each component of the loss function is evaluated through quantum subroutines with the measurement grouping.
    }
\label{fig:overview}
\end{figure}


\section{Preliminaries and Related Works}
\label{sec:Related Works}

\subsection{Spectral Operator Learning}

We begin by considering a general partial differential equation defined on a bounded $d$-dimensional domain $\Omega \subset \mathbb{R}^d$, which is characterized by a diffusion coefficient $\epsilon$, a coefficient function $a(\mathrm{x})$, and an external forcing term $f(\mathrm{x})$:
\begin{equation}
\begin{aligned}
    -\epsilon \, \text{div}(a \nabla u)+\mathcal{F}(u)
    &= f
    ,\quad
    \text{in }\Omega
    \\
    \mathcal{B}(u)&=0
    , \quad \text{on } \partial \Omega
\end{aligned}
\label{eq:reaction_diffusion}
\end{equation}
where a spatial derivative operator $\mathcal{F}$ can be either linear or nonlinear, and $\mathcal{B}$ represents a boundary operator. This formulation applies generally to second-order elliptic PDEs, including important models such as reaction--diffusion, Helmholtz and convection--diffusion equations. Within the Legendre--Galerkin framework, the governing PDE yields its weak formulation, characterized by the bilinear form $B(\cdot, \cdot)$ and the linear functional $l(\cdot)$:
\begin{equation}
\label{eq:weak_form}
    B(u,\phi):=
    \epsilon \int_\Omega a(\mathrm{x}) (\nabla u \cdot \nabla \phi)(\mathrm{x}) \, d\mathrm{x}
    + \int_\Omega [\mathcal{F}(u)(\mathrm{x})]\, \phi(\mathrm{x})\, dx
    = \int_\Omega f(\mathrm{x})\, \phi(\mathrm{x})\, d\mathrm{x}
    =: l(\phi).
\end{equation}
In the domain $I = [-1,1] \subset \mathbb{R}$, we approximate the PDE solution $u(x)$ by
\begin{equation}
    u(x) \approx \hat{u}(x) = \sum_{k=0}^{N-1} \hat{\alpha}_k \phi_k(x),
\label{eq:solution}  
\end{equation}
where the basis functions $\phi_k = L_k + a_k L_{k+1} + b_k L_{k+2}$ are defined by a compact linear combination of Legendre polynomials $\{L_k\}$. The basis coefficients $a_k$ and $b_k$ are chosen to strongly enforce the exact boundary conditions, including Dirichlet, Neumann, and mixed types. This basis representation effectively reduces the weak form \eqref{eq:weak_form} to a discrete form
\begin{equation}
\label{eq:linear_system}
    A \alpha=F,\quad\text{where}\quad
    A_{kj}=B(\phi_j,\phi_k)
    ,\quad F_k := l(\phi_k),
\end{equation}
where the spectral matrix $A$ and the forward transform $F$ are computed from the explicit calculation of the bilinear and linear forms. More generally, since the $d$-dimensional basis is defined by the tensor product of 1D basis functions, the spectral coefficient vector $\alpha$ lies in $\mathbb{R}^{N^d}$ and the matrix $A$ is in $\mathbb{R}^{N^d \times N^d}$. A detailed explanation of the weak form and matrix structures is provided in Appendix~\ref{sec:weak_form}.

Inspired by the weak formulation, the Spectral Coefficient Learning via Operator Network (SCLON)~\cite{cmame_choi2024} learns a coefficient operator that maps a forcing function to spectral coefficients of the corresponding PDE solution. To this end, SCLON trains a neural network to map PDE instances (e.g., forcing functions, initial conditions, or PDE coefficients)
to their corresponding spectral coefficients $\hat{\alpha}$ 
by minimizing an unsupervised loss function based on PDE residuals in the coefficient space.
However, as the dimensions of both input features and spectral coefficients scale as $\mathcal{O}(N^d)$, the network size suffers from the curse of dimensionality. This exponential growth with $d$ leads to prohibitive memory bottlenecks and computational costs, particularly for high-dimensional PDEs or high-accuracy simulations requiring a large number of basis modes.

\subsection{VQLS}
\label{sec:VQLS}
As a heuristic alternative to the HHL algorithm, VQLS~\cite{bravo2023} was proposed to solve linear systems with parameterized quantum circuits and classical optimization. Given a linear system $Ax=b$ of dimension $N=2^n$, VQLS aims to produce an $n$-qubit quantum state $\ket{x}$ such that $A\ket{x}$ is proportional to the target state $\ket{b}$. On the Hilbert space of $n$ qubits, the solution is represented by a parameterized quantum circuit $\ket{x} = V(\theta)\ket{0}$, while the target state is assumed to be prepared as $\ket{b}=U\ket{0}$ for a fixed unitary operator $U$. The matrix $A$ is expressed as a linear combination of unitaries,
\begin{equation}
    A = \sum_{l}^{L} c_{l} A_{l},
\end{equation}
where $A_l \in \mathbb{C}^{2^n \times 2^n}$ are typically Pauli operators, or more generally unitary basis elements, and $c_l$ are complex coefficients. The standard VQLS objective maximizes the fidelity between the normalized state
$A\ket{x}/\sqrt{\bra{x}A^\dagger A\ket{x}}$ and the target state $\ket{b}$, which can be written as
\begin{equation}
\label{eq:VQLS cost function}
\begin{aligned}
    \mathcal{L}_{\mathrm{VQLS}}(\theta) 
    =
    1 -     
    \frac{
    \sum_{l,l'}^{L} c_{l'}^*c_l
    \langle 0  \lvert
    V(\theta)^\dagger A_{l'}^\dagger U\ket{0} \bra{0} U^\dagger A_l  V(\theta) \lvert 0 \rangle}
    {
    \sum_{l,l'}^{L} c_{l'}^*c_l
    \langle 0  \lvert V^\dagger(\theta)  A_{l'}^\dagger A_l V(\theta) \lvert 0 \rangle} \, .  
\end{aligned}
\end{equation}
Equivalently, this objective can be viewed as minimizing the expectation value of the effective Hamiltonian $H \equiv A^\dagger (I - \ket{b}\bra{b}) A$. This energy-based formulation is useful for variational optimization, but it also makes the objective insensitive to global sign or phase information of the solution state. This becomes problematic in PDE settings, where the sign of the solution carries physical and numerical meaning. Because of the double-sum structure in Eq.~\eqref{eq:VQLS cost function}, standard VQLS requires $\mathcal{O}(L^2)$ expectation evaluations, often implemented through controlled-unitary operations, which can lead to substantial circuit-depth overhead.

\begin{figure}[t]
    \begin{center}
    \includegraphics[width=\linewidth]{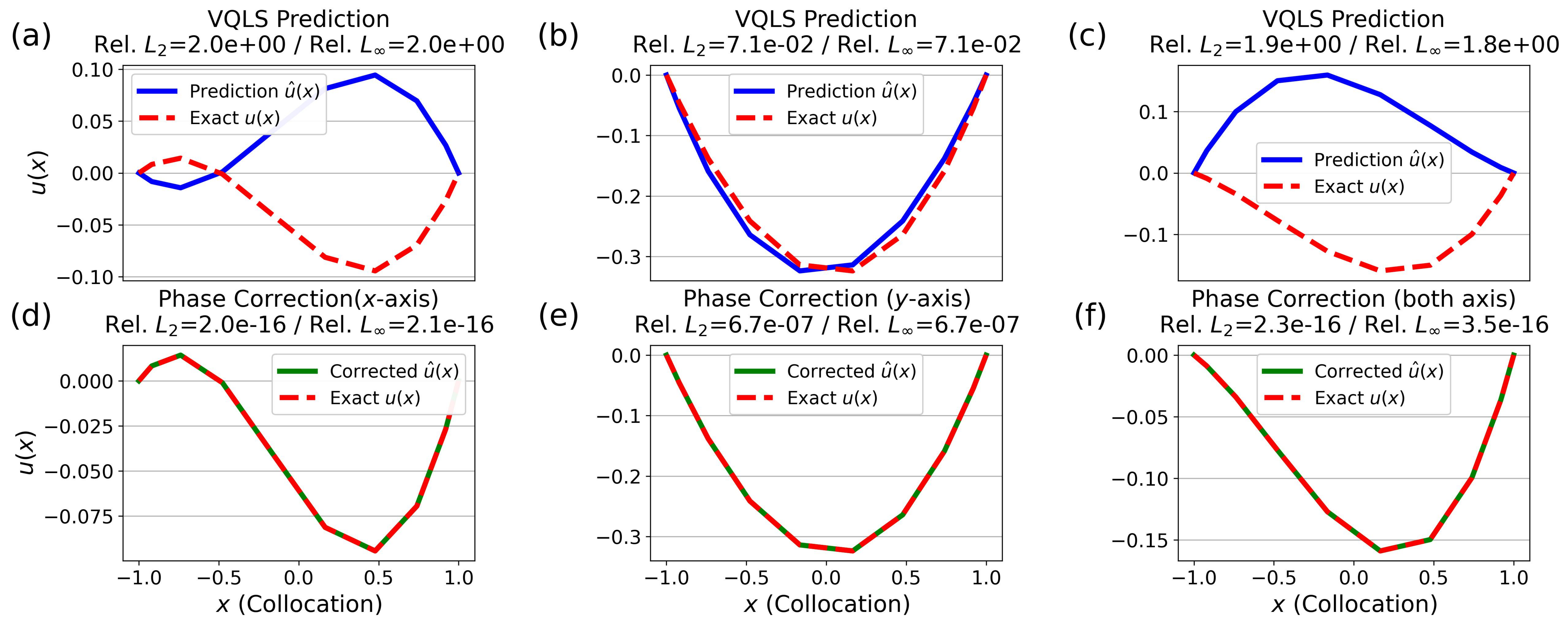}
    \end{center}
    \caption{
Illustration of the reflection error inherent in standard VQLS and the effect of phase correction. 
Rows from top to bottom display (a--c) the initial VQLS predictions exhibiting reflection errors, and (d--f) the corresponding solutions after applying the phase correction. 
Columns from left to right correspond to reflections along the $x$-axis, the $y$-axis, and both axes, respectively.
} \label{fig:reflection_Helm}
\end{figure}

\subsection{Related Works} 

Previous applications of VQLS for solving PDEs have largely been confined to problems yielding highly structured matrices \cite{liu2021variational,liu2022application,e25040580,ali2023performance}. Other works have focused on improving the practicality of VQLS and broadening its scope of applications~\cite{turati2024empirical,pellow2023near,patil2022variational,pellow2021comparison,surana2024variational,hosaka2023preconditioning,gnanasekaran2024efficient}. Yet, most VQLS-based approaches are typically restricted to instance-specific training, requiring retraining for every new problem setup. 
In parallel, quantum approaches for operator learning, such as QFNO~\cite{jain2024quantum} and Quantum DeepONet~\cite{xiao2025quantum}, use variational quantum circuits as function approximators for PDE solutions. However, these methods follow a supervised learning paradigm and require large collections of precomputed solution data. In contrast, our work develops a hybrid quantum-classical framework for unsupervised operator learning, where PDE structure is incorporated into the training objective, predicting spectral solution coefficients across problem instances without labeled solutions.

\begin{figure}[t]
    \begin{center}
    \includegraphics[width=\linewidth]{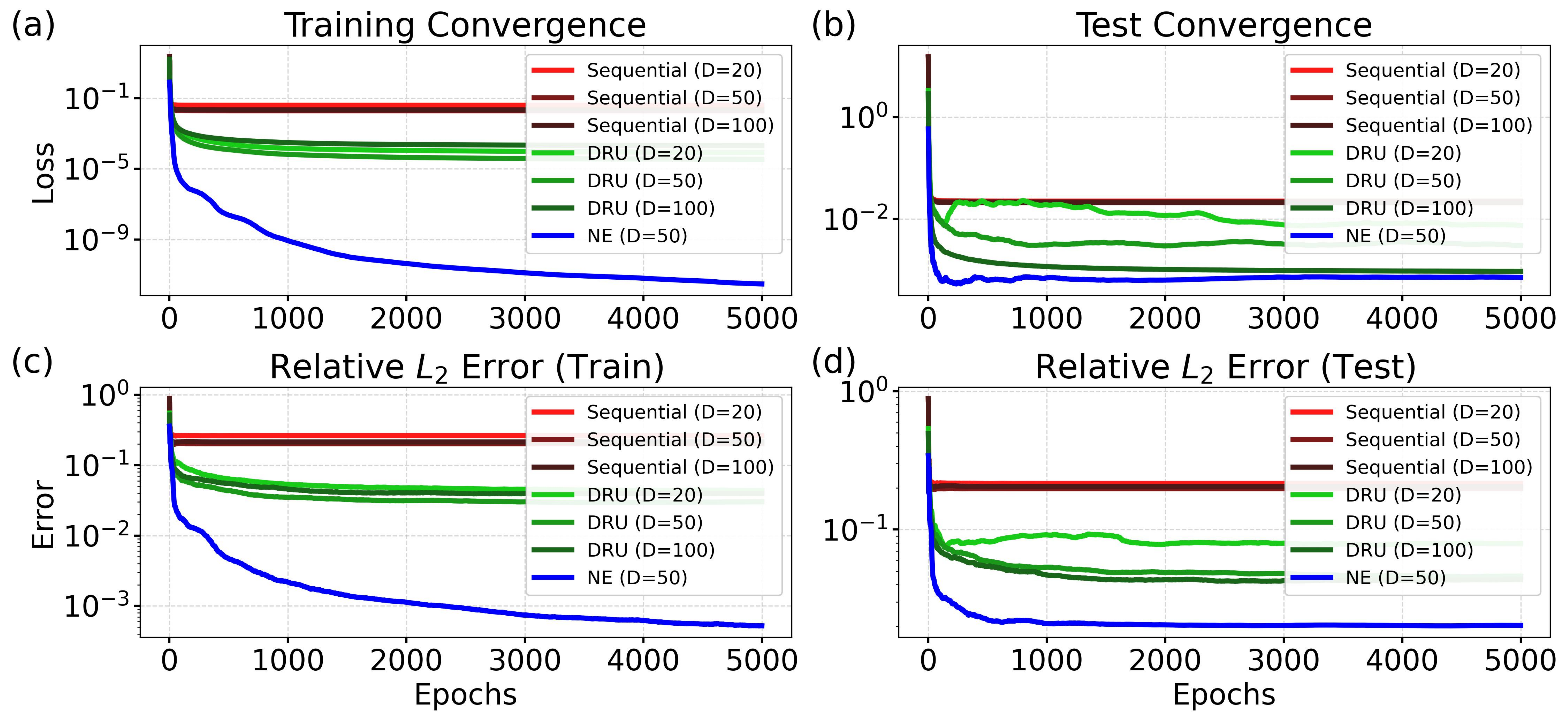}
    \end{center}
    \caption{
    Ablation study on the effect of Neural Embedding (NE). Plots (a--d) show convergence histories for training/test losses and relative $L_2$ errors. We compare NVQLS with NE (7,780 parameters, depth 10) using $D=50$ against two quantum baselines (1,200 parameters, depth 100) evaluated across $D \in \{20, 50, 100\}$. Here, $D$ denotes the number of training instances. The results highlight that neural embedding enables effective multi-instance learning with a significantly shallower circuit.
    } \label{fig:ablation_neural_embedding}
\end{figure}

\section{Methodology}
\label{sec:Methodology}

In this section, we introduce the Neural Variational Quantum Linear Solver (NVQLS), a hybrid variational quantum algorithm designed to approximate the spectral operator that maps a batch of forcing functions $\{f^{(i)}\}_{i=1}^D$ to their corresponding solution coefficients. The training objective of NVQLS is to minimize weak-form residuals in the spectral coefficient space. Figure~\ref{fig:overview} provides an overview of the NVQLS training procedure. 

\subsection{Phase-Aware Overlap Cost Function}

The standard VQLS objective determines the solution state only up to a global phase. 
Consequently, when applied to real-valued PDE solution coefficients, it can lead to sign or reflection errors in the predicted solution, as illustrated in Fig.~\ref{fig:reflection_Helm}. To resolve this ambiguity, we introduce a \textit{phase-aware overlap cost} that directly uses the real part of the quantum state overlap to recover the correct solution sign:

\begin{equation}
\label{eq:loss_NVQLS}
\begin{aligned}
    \mathcal{L}_{\mathrm{NVQLS}}(w)
    =\frac{1}{D}\sum_{i=1}^{D} \left( 1 - \frac{
    \mathrm{Re}\left( \sum_{l=1}^{L}
    c_{l} 
    \bra{ F^{(i)} } A_{l} 
    \ket{ {\hat{\alpha}(f^{(i)};w) } } \right)}
    {\sqrt{
    \sum_{l=1}^{R} d_{l}
    \bra{\hat{\alpha}(f^{(i)};w)}
    A_{l}
    \ket{\hat{\alpha}(f^{(i)};w)}}
    }\right),
\end{aligned}
\end{equation}
where the system operator and its normal operator are decomposed as
$A =\sum_{l=1}^{L} c_{l} A_{l}$ and
$A^\dagger A = \sum_{l=1}^{R} d_{l} A_{l}$
with an $n$-qubit Pauli operator $A_l\in \lbrace I,X,Y,Z\rbrace^{\otimes n}$. Here, the forward transform of each forcing function is represented by unitaries $\ket{F^{(i)}}=U^{(i)}\ket{0}$.
This formulation avoids the double-sum structure of the standard VQLS numerator by using a linear overlap between the transformed forcing vector and the predicted residual vector.
Moreover, the Pauli operators in Eq.~\eqref{eq:loss_NVQLS} appear as observables measured at the end of the circuit, rather than as controlled unitaries inside Hadamard-test-type circuits. This structure enables commuting Pauli strings to be grouped and measured together, reducing the number of circuit evaluations required for loss estimation. A detailed derivation of Eq.~\eqref{eq:loss_NVQLS} is provided in Appendix~\ref{sec:loss_derivation}. 

\subsection{Neural Embedding for Multiple Instance Learning}
\label{sec:neural_embedding}

To enable generalization across multiple PDE instances, we introduce a \textit{neural embedding} scheme that maps instance-dependent PDE inputs to quantum circuit parameters. Specifically, a classical neural network $g$ with trainable weights $w$ takes each forcing function $f^{(i)}$ as input and outputs the corresponding circuit parameters $\theta^{(i)}=g(f^{(i)};w)$. The predicted solution coefficients are then encoded in the quantum state
\begin{equation}
    \ket{\hat{\alpha}}=
     V(\theta^{(i)}) \ket{0} = V\left(g (f^{(i)};w)\right)\ket{0}.
\label{eq:variational_ansatz}
\end{equation}

To validate the proposed hybrid structure, Fig.~\ref{fig:ablation_neural_embedding} presents an ablation study comparing our quantum-classical neural embedding against two purely quantum baselines: a sequential embedding formulated as $V(\theta, f) = V(\theta)U(f)$ and a data re-uploading scheme, which interleaves data encoding layers throughout the variational circuit to increase expressivity. Both baselines use 100 variational layers with approximately 1,200 parameters, but exhibit training difficulties as the number of instances increases ($D=20, 50, \text{ and } 100$). In contrast, NVQLS with neural embedding generalizes to 200 test samples from a training set of only $D=50$, using 10 layers with 7,780 parameters. As further detailed in Appendix~\ref{sec:ablation_neural_embedding} and Fig.~\ref{fig:efficient_ablation}, neural embedding achieves superior convergence and generalization even with only 2 variational layers and 1,304 parameters.

Furthermore, by adopting this hybrid architecture, the neural network projects the input forcing function onto a compressed parameter space of dimension $\mathcal{O}(n)=\mathcal{O}(\log^2 N^d)$. This reduced output dimension lowers the computational cost of both forward and backward passes through the classical network.

\begin{table}[t]
\caption{
Per-iteration time and memory complexities as functions of the system size $N^d$.
The time complexity is reported for a target observable-estimation accuracy $\epsilon$.
}
\label{tab:complexity}
\begin{center}
\begin{tabular}{lcc}
    \toprule
    Model
    & NVQLS
    & \makecell{SCLON / PI-DON} \\
    \midrule
    Time Complexities
        & $\mathcal{O}\!\left({N^d(\log N^d)^4}/{\epsilon^2}\right)$
        & $\mathcal{O}\!\left(N^{2d}\right)$ \\
    Memory Complexities
    & $\mathcal{O}\!\left({N^d(\log N^d)^2}\right)$
    & $\mathcal{O}\!\left(N^{2d}\right)$ \\
    
    \bottomrule
\end{tabular}
\end{center}
\end{table}

\subsection{Complexity Analysis with Efficient Circuit Implementation}
\label{sec:complexity_analysis}

We analyze the per-iteration runtime and memory complexities of NVQLS and summarize the resulting upper bounds in Table~\ref{tab:complexity}, together with those of classical operator-learning methods. NVQLS improves the runtime and memory scaling over existing classical neural operator methods. This improvement comes from two key features of the proposed framework: the measurement-grouping strategy enabled by the circuit structure and the compact hybrid parameterization, which reduces the number of trainable parameters. 

\begin{proposition}[Per-Iteration Runtime Complexity of NVQLS]
\label{prop:runtime_complexity}
Assume that the $f$-embedding and variational circuits admit efficient implementations whose depths scale polylogarithmically with the problem size $N^d$. Then, for a target observable-estimation accuracy $\epsilon>0$, each NVQLS training iteration requires at most
$
    \mathcal{O}\!\left(
    {N^d(\log N^d)^4}/{\epsilon^2}
    \right)
$
time.
If the truncated Pauli approximation is applied, this per-iteration bound is further reduced to
$
    \mathcal{O}\!\left(
    N^d(\log N^d)^2
    \right).
$
\begin{proof}
See Appendix~\ref{sec:detailed_complexity_analysis} for the detailed derivation. 
\end{proof}
\end{proposition}

\begin{proposition}[Per-Iteration Memory Complexity of NVQLS]
\label{prop:memory_complexity}
Under the same assumptions as Proposition~\ref{prop:runtime_complexity}, the
per-iteration classical working-memory complexity of NVQLS is in
$
    \mathcal{O}\!\left(
    N^d(\log N^d)^2
    \right).
$
The same upper bound holds when the truncated Pauli approximation is applied.
\begin{proof}
See Appendix~\ref{sec:detailed_complexity_analysis} for the detailed derivation.
\end{proof}
\end{proposition}

For classical unsupervised neural-operator methods, the per-iteration time
complexity is dominated by neural-network forward and backward passes, typically
scaling as $\mathcal{O}(N^{2d})$ arithmetic operations. Under the assumptions detailed in Appendix~\ref{sec:detailed_complexity_analysis}, the
per-iteration memory complexity scales with the number of trainable parameters,
which is in $\mathcal{O}(N^{2d})$.

As a quantum baseline, standard VQLS solves each input instance separately,
requiring total runtime
$D \,T_{\mathrm{iter}} \, T_{\mathrm{per\text{-}iter}}$ for $D$ instances,
where $T_{\mathrm{iter}}$ is the number of optimization iterations and
$T_{\mathrm{per\text{-}iter}}$ denotes the per-iteration runtime cost.
Under comparable circuit-depth and gradient-estimation assumptions, the
per-iteration time complexity of standard VQLS is in
$
    \mathcal{O}\!\left(
    {N^{2d}(\log N^d)^5}/{\epsilon^2}
    \right).
$ 
In contrast, NVQLS generalizes across multiple input instances while reducing
the per-iteration time complexity relative to this quantum baseline.

\section{Experimental Results}
\label{sec:Experiment}

\subsection{Spectral Operator Learning with External Forcing Inputs}
\label{sec:1d_pdes}

In this section, we describe numerical results for various 1D equations to assess the numerical accuracy of our model. Following the generation of forcing functions under shallow embedding and general settings (Table~\ref{tab:instance_generation}), NVQLS maps these inputs to their corresponding spectral coefficients. Detailed training setups and network architectures are deferred to Appendix~\ref{sec:training_details}, while problem formulations are provided in Appendix~\ref{sec:experiment_details}.

\begin{figure}[t]
\begin{center}
\includegraphics[width=\linewidth]{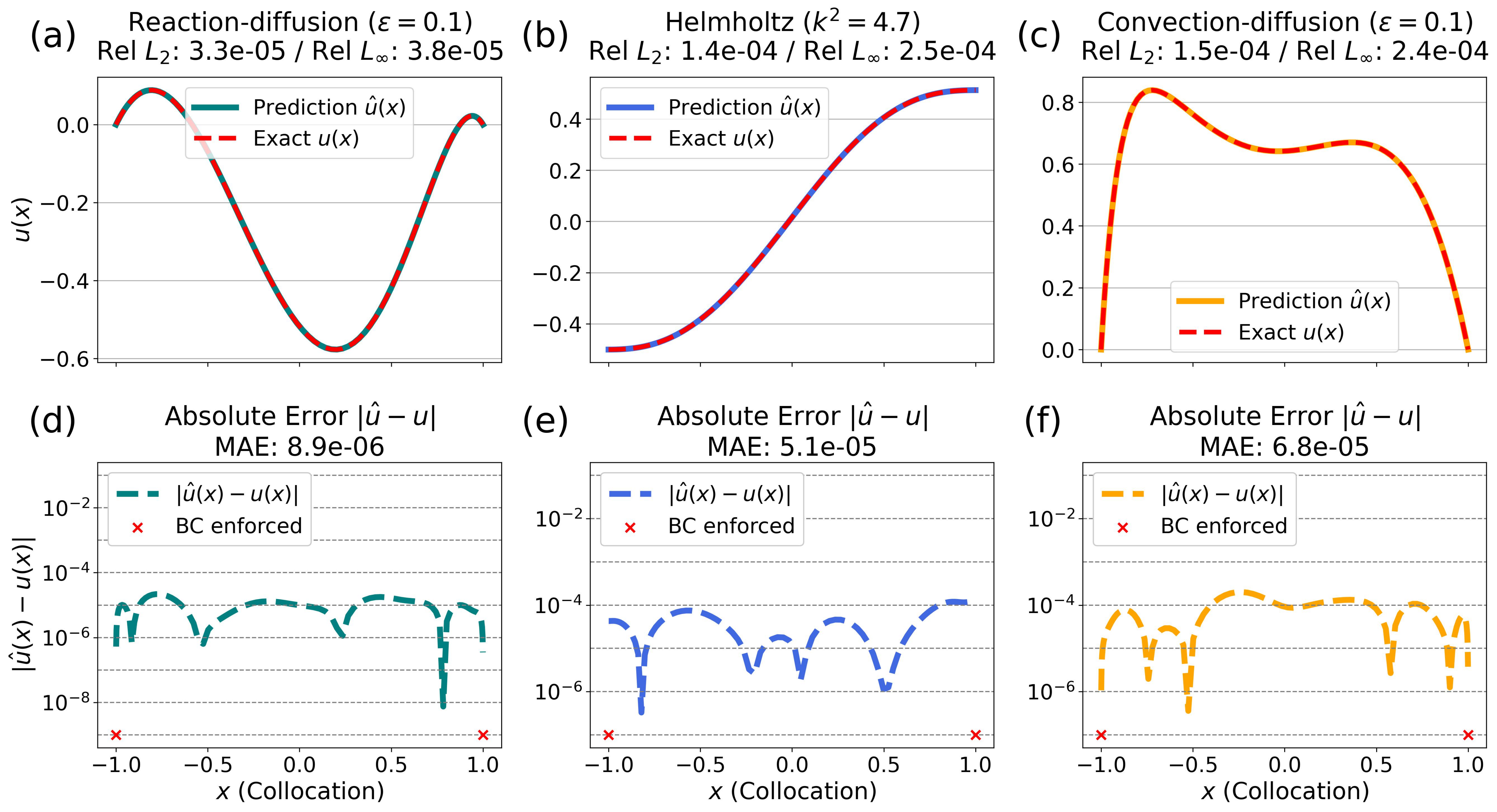}
\end{center}
    \caption{
    Numerical results for 1D steady-state elliptic PDEs. 
    Top row: (a--c) Predicted solution $\hat{u}$ versus exact solution $u$. 
    Bottom row: (d--f) Absolute error $|\hat{u} - u|$. 
    Columns from left to right correspond to the reaction--diffusion ($\epsilon=0.1$, Dirichlet BC), Helmholtz ($k=4.7$, Neumann BC), and convection--diffusion ($\epsilon=0.1$, Dirichlet BC) equations, respectively. 
    } \label{fig:1d_pdes_example}
\end{figure}

\textbf{1D Steady-state Elliptic PDEs.}
Figure~\ref{fig:1d_pdes_example} summarizes the numerical results for one-dimensional PDE benchmarks. The visualized plots represent the best-performing instances, achieving the lowest relative $L^2$ error among 200 test samples. These experiments on various boundary conditions, including the homogeneous Dirichlet and Neumann boundary conditions, highlight the robustness of NVQLS across different PDE constraints. 
Detailed statistical information regarding the test errors is summarized in Table~\ref{tab:error_table}. Across all 1D and 2D PDE benchmarks, the mean relative $L_2$ and $L_\infty$ errors are consistently maintained at low orders of magnitude, ranging from $10^{-2}$ to $10^{-3}$. Notably, for several cases such as 1D reaction--diffusion and the joint Helmholtz problem, the average errors reach as low as approximately $0.1\%$. These results demonstrate the robust generalization capability of the NVQLS framework across diverse physical systems.

\begin{figure}[t]
\begin{center}
    \includegraphics[width=\linewidth]{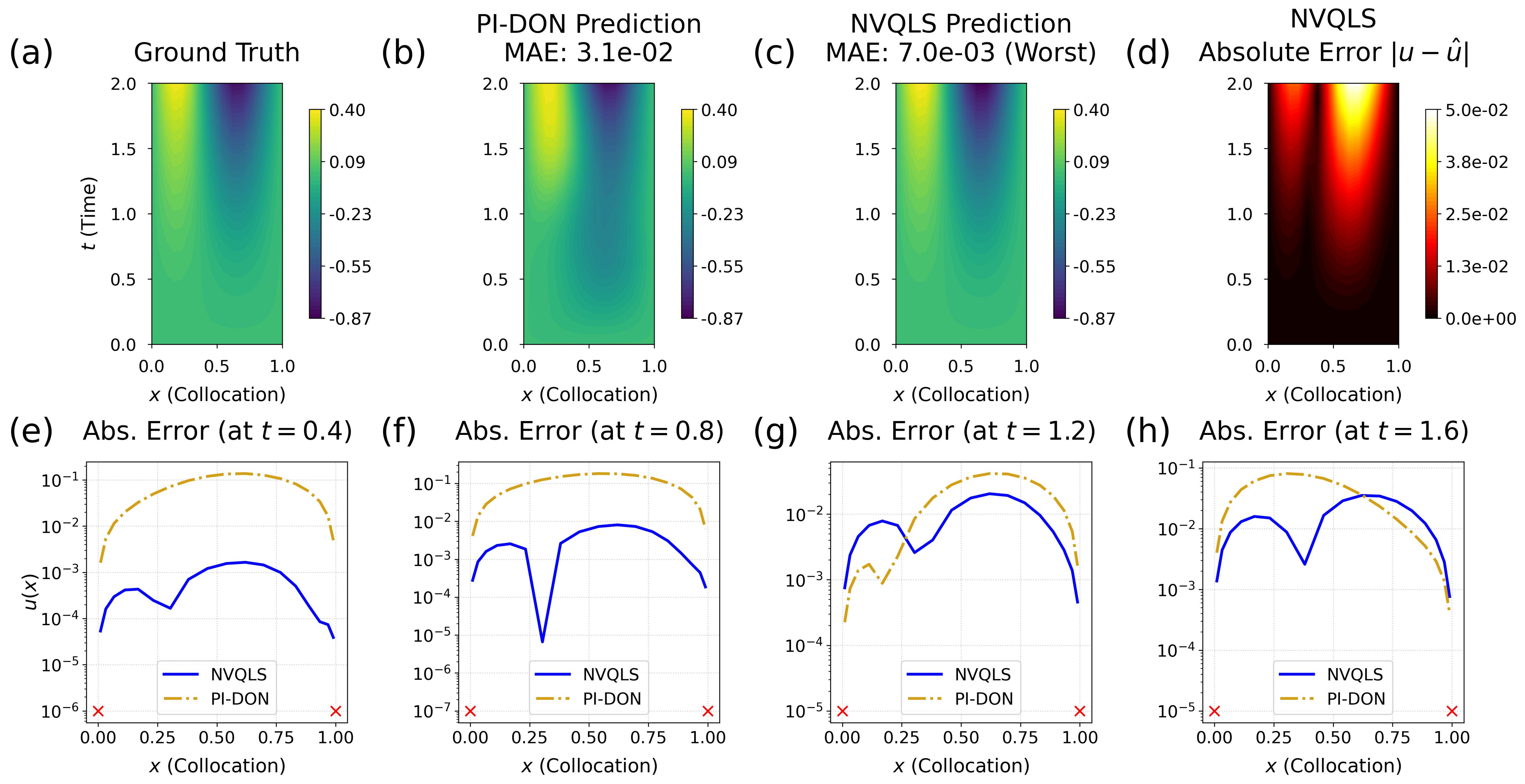}
\end{center}
    \caption{Representative solution profiles for the wave equation. Row 1: Spatio-temporal profiles of (a) the ground truth, (b) the PI-DON prediction, (c) the NVQLS prediction, and (d) the absolute error of the NVQLS prediction. Row 2: (e--h) Snapshots comparing the absolute errors of each model at $t=0.4, 0.8, 1.2$, and $1.6$. Notably, across 200 test samples, even the worst-case prediction of NVQLS demonstrates higher accuracy than the PI-DON prediction.}
\label{fig:comparison_wave_sol_profile}
\end{figure}

\textbf{Classical Baseline Comparison on 1D Wave Equation.} To evaluate the model's capability to capture complex temporal dynamics, we trained our proposed NVQLS on the time-dependent, hyperbolic wave equation. We then compared its performance against a classical physics-informed deep operator network (PI-DON), a representative unsupervised operator learning framework. Figure~\ref{fig:comparison_wave_sol_profile} presents a comprehensive comparison of the solution profiles, including the ground truth, predictions from both models, and the absolute error of NVQLS. Time snapshots at $t=0.4, 0.8, 1.2,$ and $1.6$ are also provided to closely examine the solution fidelity. To ensure a rigorous comparison, we specifically visualized the worst-performing sample for NVQLS with the maximum mean absolute error (MAE) among 200 test cases. Remarkably, even in this worst-case scenario, NVQLS outperforms the PI-DON prediction, maintaining an exceptionally low error magnitude on the order of $10^{-3}$. This demonstrates its superior capability to resolve complex wave dynamics more accurately than the classical PI-DON baseline.

\subsection{Extension to Two-dimensional Parametric PDEs for Joint Coefficient–Forcing Operator Learning}
\label{sec:joint_2d_Helm}

We now extend spectral operator learning to two-dimensional PDE problems, where the operator maps jointly varying coefficient and forcing function inputs to their corresponding solutions. For training, the input instances are tuples of a forcing function $f^{(i)}$ and PDE coefficients $k^{(i)}$, represented as $\{ (k^{(i)}, f^{(i)}) \}_{i=1}^D$. Each tuple is mapped to quantum circuit parameters $\theta^{(i)} = g(k^{(i)}, f^{(i)}; w)$ via the angle network $g$. We demonstrate this using the two-dimensional Helmholtz equation, characterized by a wave number $k^2$:
\begin{equation}
\begin{aligned}
    \Delta u(x,y)
    + k^2 u(x,y)
    &= f(x,y)
    ,\quad \text{in} \, \Omega
    \\
    \mathcal{B}(u)(x,y) &=0
    ,\quad \text{on} \, \partial \Omega
\end{aligned}
\end{equation}
Since the corresponding spectral method matrix is determined by the parameter $k$ and involves tensor products of fixed one-dimensional matrices, we separate the constant and parameter-dependent terms of the matrix (i.e., $B + k^2C$) in optimization, enabling us to execute the Pauli decomposition just twice rather than for every instance of $k$. 

\begin{figure}[t]
\begin{center}
    \includegraphics[width=\linewidth]{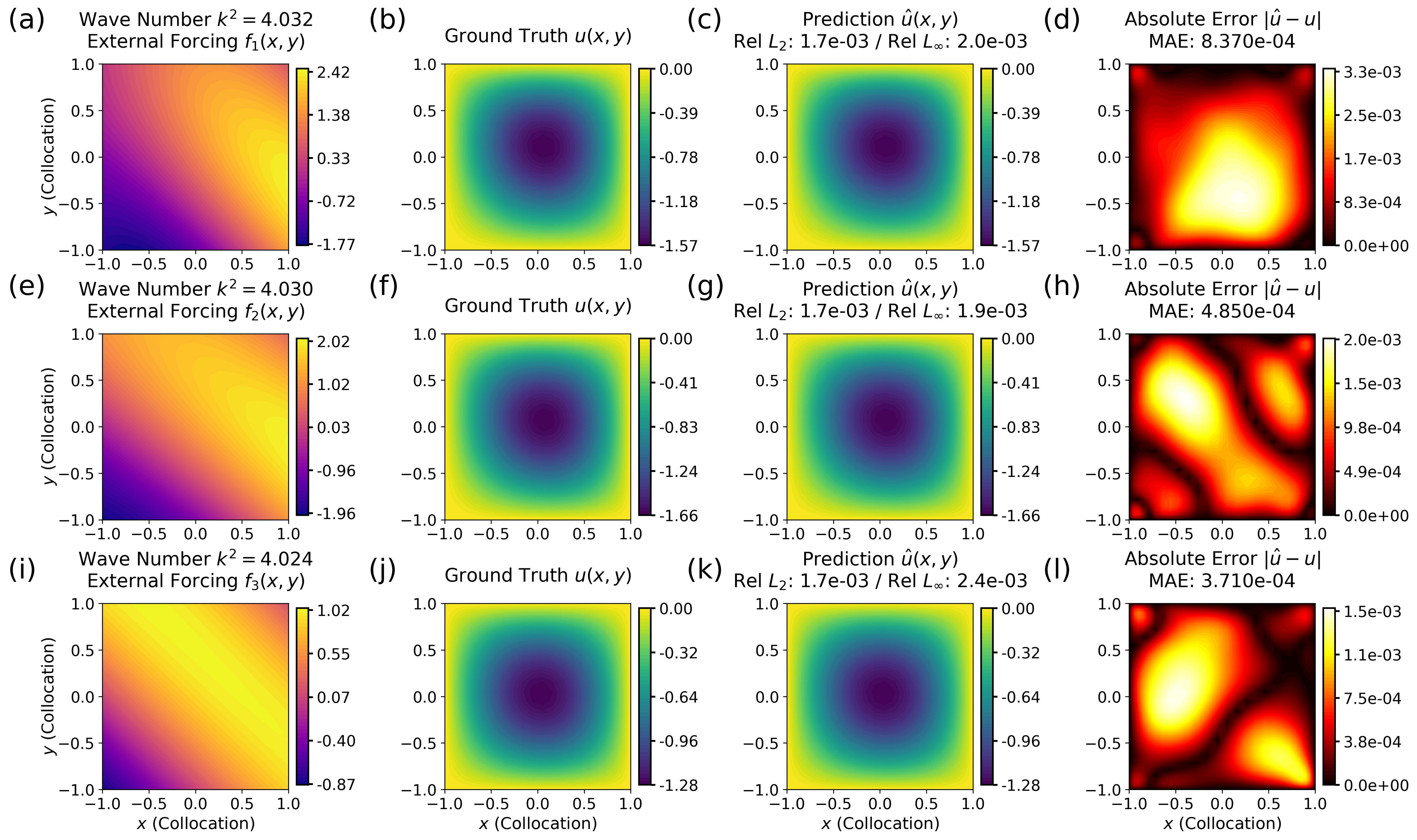}
\end{center}
    \caption{Numerical examples of operator learning with joint parameter and forcing inputs for the two-dimensional Helmholtz equation (Dirichlet BC).
    Row 1 (Case 1): (a) Input pairs $f_1$ and $k^2=4.032$ for the angle network, (b) ground truth $u_1$, (c) NVQLS prediction $\hat{u}_1$, and (d) absolute error $|\hat{u}_1 - u_1|$ on interior nodal points. Row 2 and Row 3 show the corresponding results for Case 2 ($f_2$ and $k^2=4.030$) and Case 3 ($f_3$ and $k^2 = 4.024$), respectively.
    }
\label{fig:joint_Helm_2d}
\end{figure}

Figure~\ref{fig:joint_Helm_2d} presents three representative examples of operator learning with joint parameter and forcing inputs for the two-dimensional Helmholtz equation, using the homogeneous Dirichlet boundary conditions. Each row corresponds to a specific test instance. From left to right, the first column presents the input instances, $\{(k_i, f_i)\}_{i=1}^3$, where the wave numbers $k_i$ are sampled from the uniform distribution and the forcing functions are linear combinations of trigonometric functions. The second and third columns display the exact solutions and NVQLS predictions, respectively. Finally, the fourth column illustrates their absolute differences along with MAE values. The predictions exhibit small relative errors and MAE values, demonstrating the model’s ability to approximate the solution operator across varying parameters.

\section{Conclusion}
\label{sec:Conclusion}

In this work, we proposed a hybrid quantum–classical framework for unsupervised spectral operator learning, designed to solve partial differential equations efficiently.
To achieve this, we integrated variational quantum circuits with an angle network architecture for batch training and introduced a phase-aware overlap cost function to resolve the sign ambiguity inherent in VQLS. 
Our theoretical and empirical analysis demonstrates that NVQLS significantly improves scalability, reducing both the measurement cost and neural network computational complexity scaling. Furthermore, the model exhibits broad generalization across diverse PDE instances and boundary conditions without retraining. 
These results establish NVQLS as a scalable and robust framework with a theoretical scaling advantage, laying the foundation for near-term quantum hardware implementations to solve complex scientific problems.


\newpage
{\small
\bibliographystyle{plain}
\bibliography{ref}

@article{HHL,
  title = {Quantum Algorithm for Linear Systems of Equations},
  author = {Harrow, Aram W. and Hassidim, Avinatan and Lloyd, Seth},
  journal = {Phys. Rev. Lett.},
  volume = {103},
  issue = {15},
  pages = {150502},
  numpages = {4},
  year = {2009},
  month = {Oct},
  publisher = {American Physical Society},
  doi = {10.1103/PhysRevLett.103.150502},
  url = {https://link.aps.org/doi/10.1103/PhysRevLett.103.150502}
}

@Article{e26080649,
AUTHOR = {Trahan, Corey and Loveland, Mark and Dent, Samuel},
TITLE = {Quantum Physics-Informed Neural Networks},
JOURNAL = {Entropy},
VOLUME = {26},
YEAR = {2024},
NUMBER = {8},
ARTICLE-NUMBER = {649},
URL = {https://www.mdpi.com/1099-4300/26/8/649},
PubMedID = {39202119},
ISSN = {1099-4300},
DOI = {10.3390/e26080649}
}

@article{PhysRevA.110.022612,
  title = {Deep-learning-based quantum algorithms for solving nonlinear partial differential equations},
  author = {Mouton, Lukas and Reiter, Florentin and Chen, Ying and Rebentrost, Patrick},
  journal = {Phys. Rev. A},
  volume = {110},
  issue = {2},
  pages = {022612},
  numpages = {33},
  year = {2024},
  month = {Aug},
  publisher = {American Physical Society},
  doi = {10.1103/PhysRevA.110.022612},
  url = {https://link.aps.org/doi/10.1103/PhysRevA.110.022612}
}

@article{Kyriienko2021,
  title = {Solving nonlinear differential equations with differentiable quantum circuits},
  author = {Kyriienko, Oleksandr and Paine, Annie E. and Elfving, Vincent E.},
  journal = {Phys. Rev. A},
  volume = {103},
  issue = {5},
  pages = {052416},
  numpages = {22},
  year = {2021},
  month = {May},
  publisher = {American Physical Society},
  doi = {10.1103/PhysRevA.103.052416},
  url = {https://link.aps.org/doi/10.1103/PhysRevA.103.052416}
}

@article{Sedykh_2024,
doi = {10.1088/2632-2153/ad43b2},
url = {https://doi.org/10.1088/2632-2153/ad43b2},
year = {2024},
month = {may},
publisher = {IOP Publishing},
volume = {5},
number = {2},
pages = {025045},
author = {Sedykh, Alexandr and Podapaka, Maninadh and Sagingalieva, Asel and Pinto, Karan and Pflitsch, Markus and Melnikov, Alexey},
title = {Hybrid quantum physics-informed neural networks for simulating computational fluid dynamics in complex shapes},
journal = {Machine Learning: Science and Technology}
}

@article{jain2024quantum,
  title={Quantum Fourier networks for solving parametric PDEs},
  author={Jain, Nishant and Landman, Jonas and Mathur, Natansh and Kerenidis, Iordanis},
  journal={Quantum Science and Technology},
  volume={9},
  number={3},
  pages={035026},
  year={2024},
  publisher={IOP Publishing}
}

@article{xiao2025quantum,
  title={Quantum DeepONet: Neural operators accelerated by quantum computing},
  author={Xiao, Pengpeng and Zheng, Muqing and Jiao, Anran and Yang, Xiu and Lu, Lu},
  journal={Quantum},
  volume={9},
  pages={1761},
  year={2025},
  publisher={Verein zur F{\"o}rderung des Open Access Publizierens in den Quantenwissenschaften}
}

@misc{morales2025,
      title={Quantum Linear System Solvers: A Survey of Algorithms and Applications}, 
      author={Mauro E. S. Morales and Lirandë Pira and Philipp Schleich and Kelvin Koor and Pedro C. S. Costa and Dong An and Alán Aspuru-Guzik and Lin Lin and Patrick Rebentrost and Dominic W. Berry},
      year={2025},
      eprint={2411.02522},
      archivePrefix={arXiv},
      primaryClass={quant-ph},
      url={https://arxiv.org/abs/2411.02522}, 
}

@article{bravo2023,
  title={Variational quantum linear solver},
  author={Bravo-Prieto, Carlos and LaRose, Ryan and Cerezo, Marco and Subasi, Yigit and Cincio, Lukasz and Coles, Patrick J},
  journal={Quantum},
  volume={7},
  pages={1188},
  year={2023},
  publisher={Verein zur F{\"o}rderung des Open Access Publizierens in den Quantenwissenschaften}
}

@article{PIDON,
    author = {Sifan Wang  and Hanwen Wang  and Paris Perdikaris },
    title = {Learning the solution operator of parametric partial differential equations with physics-informed DeepONets},
    journal = {Science Advances},
    volume = {7},
    number = {40},
    pages = {eabi8605},
    year = {2021},
    doi = {10.1126/sciadv.abi8605},
    URL = {https://www.science.org/doi/abs/10.1126/sciadv.abi8605},
    eprint = {https://www.science.org/doi/pdf/10.1126/sciadv.abi8605}
}

@article{li2020,
  author    = {Zongyi Li and Nikola B. Kovachki and Kamyar Azizzadenesheli and Burigede Liu and Kaushik Bhattacharya and Andrew M. Stuart and Anima Anandkumar},
  title     = {Fourier Neural Operator for Parametric Partial Differential Equations},
  journal   = {arXiv preprint arXiv:2010.08895},
  year      = {2020}
}

@article{li2021,
  author    = {Zongyi Li and Hongkai Zheng and Nikola B. Kovachki and David Jin and Haoxuan Chen and Burigede Liu and Kamyar Azizzadenesheli and Anima Anandkumar},
  title     = {Physics-Informed Neural Operator for Learning Partial Differential Equations},
  journal   = {arXiv preprint arXiv:2111.03794},
  year      = {2021}
}

@article{lu2021learning,
  title={Learning nonlinear operators via DeepONet based on the universal approximation theorem of operators},
  author={Lu, Lu and Jin, Pengzhan and Pang, Guofei and Zhang, Zhongqiang and Karniadakis, George Em},
  journal={Nature machine intelligence},
  volume={3},
  number={3},
  pages={218--229},
  year={2021},
  publisher={Nature Publishing Group UK London}
}

@article{cmame_choi2024,
  title={Spectral operator learning for parametric PDEs without data reliance},
  author={Choi, Junho and Yun, Taehyun and Kim, Namjung and Hong, Youngjoon},
  journal={Computer Methods in Applied Mechanics and Engineering},
  volume={420},
  pages={116678},
  year={2024},
  publisher={Elsevier}
}

@article{bravo2019variational,
  title={Variational quantum linear solver},
  author={Bravo-Prieto, Carlos and LaRose, Ryan and Cerezo, Marco and Subasi, Yigit and Cincio, Lukasz and Coles, Patrick J},
  journal={arXiv preprint arXiv:1909.05820},
  year={2019}
}

@article{liu2021variational,
  title={Variational quantum algorithm for the Poisson equation},
  author={Liu, Hai-Ling and Wu, Yu-Sen and Wan, Lin-Chun and Pan, Shi-Jie and Qin, Su-Juan and Gao, Fei and Wen, Qiao-Yan},
  journal={Physical Review A},
  volume={104},
  number={2},
  pages={022418},
  year={2021},
  publisher={APS}
}

@Article{e25040580,
AUTHOR = {Trahan, Corey Jason and Loveland, Mark and Davis, Noah and Ellison, Elizabeth},
TITLE = {A Variational Quantum Linear Solver Application to Discrete Finite-Element Methods},
JOURNAL = {Entropy},
VOLUME = {25},
YEAR = {2023},
NUMBER = {4},
ARTICLE-NUMBER = {580},
URL = {https://www.mdpi.com/1099-4300/25/4/580},
PubMedID = {37190367},
ISSN = {1099-4300},
DOI = {10.3390/e25040580}
}

@article{liu2022application,
  title={Application of a variational hybrid quantum-classical algorithm to heat conduction equation and analysis of time complexity},
  author={Liu, YY and Chen, Zhen and Shu, Chang and Chew, Siou Chye and Khoo, Boo Cheong and Zhao, Xiang and Cui, YD},
  journal={Physics of Fluids},
  volume={34},
  number={11},
  year={2022},
  publisher={AIP Publishing}
}

@article{ali2023performance,
  title={Performance study of variational quantum algorithms for solving the Poisson equation on a quantum computer},
  author={Ali, Mazen and Kabel, Matthias},
  journal={Physical Review Applied},
  volume={20},
  number={1},
  pages={014054},
  year={2023},
  publisher={APS}
}

@article{turati2024empirical,
  title={An Empirical Analysis on the Effectiveness of the Variational Quantum Linear Solver},
  author={Turati, Gloria and Marruzzo, Alessia and Dacrema, Maurizio Ferrari and Cremonesi, Paolo},
  journal={arXiv preprint arXiv:2409.06339},
  year={2024}
}

@article{pellow2023near,
  title={Near term algorithms for linear systems of equations},
  author={Pellow-Jarman, Aidan and Sinayskiy, Ilya and Pillay, Anban and Petruccione, Francesco},
  journal={Quantum Information Processing},
  volume={22},
  number={6},
  pages={258},
  year={2023},
  publisher={Springer}
}

@article{patil2022variational,
  title={Variational quantum linear solver with a dynamic ansatz},
  author={Patil, Hrushikesh and Wang, Yulun and Krsti{\'c}, Predrag S},
  journal={Physical Review A},
  volume={105},
  number={1},
  pages={012423},
  year={2022},
  publisher={APS}
}

@article{pellow2021comparison,
  title={A comparison of various classical optimizers for a variational quantum linear solver},
  author={Pellow-Jarman, Aidan and Sinayskiy, Ilya and Pillay, Anban and Petruccione, Francesco},
  journal={Quantum Information Processing},
  volume={20},
  number={6},
  pages={202},
  year={2021},
  publisher={Springer}
}

@article{surana2024variational,
  title={Variational quantum framework for partial differential equation constrained optimization},
  author={Surana, Amit and Gnanasekaran, Abeynaya},
  journal={ACM Transactions on Quantum Computing},
  year={2024},
  publisher={ACM New York, NY}
}

@article{hosaka2023preconditioning,
  title={Preconditioning for a variational quantum linear solver},
  author={Hosaka, Aruto and Yanagisawa, Koichi and Koshikawa, Shota and Kudo, Isamu and Alifu, Xiafukaiti and Yoshida, Tsuyoshi},
  journal={arXiv preprint arXiv:2312.15657},
  year={2023}
}

@inproceedings{gnanasekaran2024efficient,
  title={Efficient variational quantum linear solver for structured sparse matrices},
  author={Gnanasekaran, Abeynaya and Surana, Amit},
  booktitle={2024 IEEE International Conference on Quantum Computing and Engineering (QCE)},
  volume={1},
  pages={199--210},
  year={2024},
  organization={IEEE}
}

@article{bergholm2020pennylane,
  title={Pennylane: Automatic differentiation of hybrid quantum-classical computations},
  author={Ville Bergholm and Josh Izaac and Maria Schuld and Christian Gogolin and M. Sohaib Alam and Shahnawaz Ahmed and Juan Miguel Arrazola and Carsten Blank and Alain Delgado and Soran Jahangiri and Keri McKiernan and Johannes Jakob Meyer and Zeyue Niu and Antal Száva and Nathan Killoran},
  journal={arXiv preprint arXiv:1811.04968},
  year={2020}
}

@article{bradbury2018jax,
  title={JAX: composable transformations of Python+ NumPy programs},
  author={Bradbury, James and Frostig, Roy and Hawkins, Peter and Johnson, Matthew James and Leary, Chris and Maclaurin, Dougal and Necula, George and Paszke, Adam and VanderPlas, Jake and Wanderman-Milne, Skye and others},
  year={2018}
}

@misc{agarap2019deeplearningusingrectified,
      title={Deep Learning using Rectified Linear Units (ReLU)}, 
      author={Abien Fred Agarap},
      year={2019},
      eprint={1803.08375},
      archivePrefix={arXiv},
      primaryClass={cs.NE},
      url={https://arxiv.org/abs/1803.08375}, 
}

@article{hendrycks2016gaussian,
  title={Gaussian error linear units (gelus)},
  author={Hendrycks, Dan and Gimpel, Kevin},
  journal={arXiv preprint arXiv:1606.08415},
  year={2016}
}

@article{kingma2017adam,
  title={Adam: A method for stochastic optimization},
  author={Kingma, Diederik P and Ba, Jimmy},
  journal={arXiv preprint arXiv:1412.6980},
  year={2017}
}

@article{Liu1989OnTL,
  title={On the limited memory BFGS method for large scale optimization},
  author={Dong C. Liu and Jorge Nocedal},
  journal={Mathematical Programming},
  year={1989},
  volume={45},
  pages={503-528},
  url={https://api.semanticscholar.org/CorpusID:5681609}
}

@book{shen2011spectral,
  title        = {Spectral Methods: Algorithms, Analysis and Applications},
  author       = {Shen, Jie and Tang, Tao and Wang, Li-Lian},
  series       = {Springer Series in Computational Mathematics},
  volume       = {41},
  year         = {2011},
  publisher    = {Springer Berlin, Heidelberg},
  doi          = {10.1007/978-3-540-71041-7},
    isbn         = {978-3-540-71040-0},
  url          = {https://doi.org/10.1007/978-3-540-71041-7},
  edition      = {1},
  pages        = {XVI+472}
}

@book{doi:10.1137/1.9780898719598,
author = {Trefethen, Lloyd N.},
title = {Spectral Methods in MATLAB},
publisher = {Society for Industrial and Applied Mathematics},
year = {2000},
doi = {10.1137/1.9780898719598},
address = {},
edition   = {},
URL = {https://epubs.siam.org/doi/abs/10.1137/1.9780898719598},
eprint = {https://epubs.siam.org/doi/pdf/10.1137/1.9780898719598}
}

@book{Burden1989,
  address = {Boston},
  author = {Burden, Richard L. and Faires, J. Douglas},
  edition = {Fourth},
  publisher = {{PWS-Kent} Publishing Company},
  series = {The Prindle, Weber and Schmidt Series in Mathematics},
  title = {Numerical Analysis},
  username = {jil},
  year = 1989
}

@article{Quantum_Spectral_Methods_for_Differential_Equations,
	author = {Childs, Andrew M. and Liu, Jin-Peng},
	date = {2020/04/01},
	doi = {10.1007/s00220-020-03699-z},
	id = {Childs2020},
	isbn = {1432-0916},
	journal = {Communications in Mathematical Physics},
	number = {2},
	pages = {1427--1457},
	title = {Quantum Spectral Methods for Differential Equations},
	url = {https://doi.org/10.1007/s00220-020-03699-z},
	volume = {375},
	year = {2020}}

@Article{make7040137,
    AUTHOR = {Zhang, Guangya and Xu, Tie and Xu, Jinli and Wang, Hu},
    TITLE = {A Graph-Structured, Physics-Informed DeepONet Neural Network for Complex Structural Analysis},
    JOURNAL = {Machine Learning and Knowledge Extraction},
    VOLUME = {7},
    YEAR = {2025},
    NUMBER = {4},
    ARTICLE-NUMBER = {137},
    URL = {https://www.mdpi.com/2504-4990/7/4/137},
    ISSN = {2504-4990},
    DOI = {10.3390/make7040137}
}

@misc{cho2025physicsinformeddeepinverseoperator,
      title={Physics-Informed Deep Inverse Operator Networks for Solving PDE Inverse Problems}, 
      author={Sung Woong Cho and Hwijae Son},
      year={2025},
      eprint={2412.03161},
      archivePrefix={arXiv},
      primaryClass={math.NA},
      url={https://arxiv.org/abs/2412.03161}, 
}

@article{LEE2026113665,
title = {Spectral coefficient learning via operator networks for inverse problems of parametric partial differential equations},
journal = {Engineering Applications of Artificial Intelligence},
volume = {166},
pages = {113665},
year = {2026},
issn = {0952-1976},
doi = {https://doi.org/10.1016/j.engappai.2025.113665},
url = {https://www.sciencedirect.com/science/article/pii/S0952197625036978},
author = {Myeong-Su Lee and Taehyun Yun and Youngjoon Hong and Namjung Kim}
}

@article{Zou22M1518189,
author = {Zou, Zongren and Meng, Xuhui and Psaros, Apostolos F. and Karniadakis, George E.},
title = {NeuralUQ: A Comprehensive Library for Uncertainty Quantification in Neural Differential Equations and Operators},
journal = {SIAM Review},
volume = {66},
number = {1},
pages = {161-190},
year = {2024},
doi = {10.1137/22M1518189},
URL = { 
        https://doi.org/10.1137/22M1518189
}
}

@misc{guo2023ibuqinformationbottleneckbased,
      title={IB-UQ: Information bottleneck based uncertainty quantification for neural function regression and neural operator learning}, 
      author={Ling Guo and Hao Wu and Wenwen Zhou and Yan Wang and Tao Zhou},
      year={2023},
      eprint={2302.03271},
      archivePrefix={arXiv},
      primaryClass={math.NA},
      url={https://arxiv.org/abs/2302.03271}, 
}

@article{10.1063_5.0188830,
    author = {Gu, Linyan and Qin, Shanlin and Xu, Lei and Chen, Ronglian },
    title = {Physics-informed neural networks with domain decomposition for the incompressible Navier–Stokes equations},
    journal = {Physics of Fluids},
    volume = {36},
    number = {2},
    pages = {021914},
    year = {2024},
    month = {02},
    issn = {1070-6631},
    doi = {10.1063/5.0188830},
    url = {https://doi.org/10.1063/5.0188830},
    eprint = {https://pubs.aip.org/aip/pof/article-pdf/doi/10.1063/5.0188830/19699194/021914_1_5.0188830.pdf},
}

@article{10.5555_3648699.3648788,
author = {Kovachki, Nikola and Li, Zongyi and Liu, Burigede and Azizzadenesheli, Kamyar and Bhattacharya, Kaushik and Stuart, Andrew and Anandkumar, Anima},
title = {Neural operator: learning maps between function spaces with applications to PDEs},
year = {2023},
issue_date = {January 2023},
publisher = {JMLR.org},
volume = {24},
number = {1},
issn = {1532-4435},
journal = {J. Mach. Learn. Res.},
month = jan,
articleno = {89},
numpages = {97}
}

@inproceedings{10.5555_3540261.3542102,
author = {Gupta, Gaurav and Xiao, Xiongye and Bogdan, Paul},
title = {Multiwavelet-based operator learning for differential equations},
year = {2021},
isbn = {9781713845393},
publisher = {Curran Associates Inc.},
address = {Red Hook, NY, USA},
booktitle = {Proceedings of the 35th International Conference on Neural Information Processing Systems},
articleno = {1841},
numpages = {15},
series = {NIPS '21}
}

@article{
doi:10.1126/sciadv.1602614,
author = {Samuel H. Rudy  and Steven L. Brunton  and Joshua L. Proctor  and J. Nathan Kutz },
title = {Data-driven discovery of partial differential equations},
journal = {Science Advances},
volume = {3},
number = {4},
pages = {e1602614},
year = {2017},
doi = {10.1126/sciadv.1602614},
URL = {https://www.science.org/doi/abs/10.1126/sciadv.1602614},
eprint = {https://www.science.org/doi/pdf/10.1126/sciadv.1602614}
}

@article{
doi:10.1126/sciadv.abk0644,
author = {Enrui Zhang  and Ming Dao  and George Em Karniadakis  and Subra Suresh },
title = {Analyses of internal structures and defects in materials using physics-informed neural networks},
journal = {Science Advances},
volume = {8},
number = {7},
pages = {eabk0644},
year = {2022},
doi = {10.1126/sciadv.abk0644},
URL = {https://www.science.org/doi/abs/10.1126/sciadv.abk0644},
eprint = {https://www.science.org/doi/pdf/10.1126/sciadv.abk0644}}

@article{Koric_s00366-023-01822-x,
	author = {Koric, Seid and Viswantah, Asha and Abueidda, Diab W. and Sobh, Nahil A. and Khan, Kamran},
	date = {2024/04/01},
	doi = {10.1007/s00366-023-01822-x},
	id = {Koric2024},
	isbn = {1435-5663},
	journal = {Engineering with Computers},
	number = {2},
	pages = {917--929},
	title = {Deep learning operator network for plastic deformation with variable loads and material properties},
	url = {https://doi.org/10.1007/s00366-023-01822-x},
	volume = {40},
	year = {2024}}

@article{10.1007/s10915-024-02700-4,
author = {Wang, Yifan and Lin, Zhongshuo and Liao, Yangfei and Liu, Haochen and Xie, Hehu},
title = {Solving High-Dimensional Partial Differential Equations Using Tensor Neural Network and A Posteriori Error Estimators},
year = {2024},
issue_date = {Dec 2024},
publisher = {Plenum Press},
address = {USA},
volume = {101},
number = {3},
issn = {0885-7474},
url = {https://doi.org/10.1007/s10915-024-02700-4},
doi = {10.1007/s10915-024-02700-4},
journal = {J. Sci. Comput.},
month = nov,
numpages = {29}
}

@article{HU2024106369,
title = {Tackling the curse of dimensionality with physics-informed neural networks},
journal = {Neural Networks},
volume = {176},
pages = {106369},
year = {2024},
issn = {0893-6080},
doi = {https://doi.org/10.1016/j.neunet.2024.106369},
url = {https://www.sciencedirect.com/science/article/pii/S0893608024002934},
author = {Zheyuan Hu and Khemraj Shukla and George Em Karniadakis and Kenji Kawaguchi}
}

@article{MENON2025118403,
title = {Anant-Net: Breaking the curse of dimensionality with scalable and interpretable neural surrogate for high-dimensional PDEs},
journal = {Computer Methods in Applied Mechanics and Engineering},
volume = {447},
pages = {118403},
year = {2025},
issn = {0045-7825},
doi = {https://doi.org/10.1016/j.cma.2025.118403},
url = {https://www.sciencedirect.com/science/article/pii/S0045782525006759},
author = {Sidharth S. Menon and Ameya D. Jagtap}
}

@InProceedings{pmlr-v242-wang24b,
  title =  {Understanding the difficulty of solving Cauchy problems with PINNs},
  author =       {Wang, Tao and Zhao, Bo and Gao, Sicun and Yu, Rose},
  booktitle = {Proceedings of the 6th Annual Learning for Dynamics \& Control Conference},
  pages = 	 {453--465},
  year = 	 {2024},
  editor = 	 {Abate, Alessandro and Cannon, Mark and Margellos, Kostas and Papachristodoulou, Antonis},
  volume = 	 {242},
  series = 	 {Proceedings of Machine Learning Research},
  month = 	 {15--17 Jul},
  publisher =    {PMLR},
  url = 	 {https://proceedings.mlr.press/v242/wang24b.html}
}

@misc{kharazmi2019variationalphysicsinformedneuralnetworks,
      title={Variational Physics-Informed Neural Networks For Solving Partial Differential Equations}, 
      author={E. Kharazmi and Z. Zhang and G. E. Karniadakis},
      year={2019},
      eprint={1912.00873},
      archivePrefix={arXiv},
      primaryClass={cs.NE},
      url={https://arxiv.org/abs/1912.00873}, 
}
}

\appendix

\newpage
\section{Nomenclature}
\label{sec:nomenclature}

\begin{table}[h]
\caption{Notations and Descriptions}
  \label{tb:nomenclature}
        \begin{tabular}{lcccr}
          \toprule
        Notation & Description \\ 
          \midrule
        $n$ 
        & Number of qubits \\
        $N$
        & Number of basis functions \\
        $d$ 
        & Dimension of PDE \\
        $K=N^d=2^n$ 
        & System Size\\
        $L$ & Number of Pauli matrices for Pauli-decomposition \\
        $x_n$
        & $n$-th nodal points on a spatial domain \\ 
        $L_k$
        & Legendre polynomial of degree $k$ \\
        $\phi_k$ 
        & $k$-th basis function
        \\ $\mathcal{F}$
        & Differential operator 
        \\ $\mathcal{B}$
        & Boundary operator 
        \\ $u(\cdot)$
        & Solution of PDE \\  
        $\alpha_k$
        & $k$-th Spectral coefficient of the solution $u(\cdot)$ \\ 
        $\hat{u}(\cdot)$
        & Approximated solution \\ 
        $\hat{\alpha}_k$
        & $k$-th Spectral coefficient of the approximated solution $\hat{u}(\cdot)$ \\
        $\hat{\alpha}$
        &  Spectral coefficient vector of the prediction $\hat{u}(\cdot)$ \\
        $f(\cdot)$ or $f^{(i)}(\cdot)$
        & ($i$-th) forcing function \\ 
        $\tilde{f}_k$ or $\tilde{f}_k^{(i)}$
        & $k$-th forward transformation of the forcing function $f$ or $f^{(i)}$ \\
        $S$ 
        & Stiffness matrix \\
        $M$ 
        & Mass matrix \\
        $R$ 
        & Convection matrix \\
        $A$ 
        & Spectral matrix corresponding to PDE problem 
        \\
        $c_l$ 
        & Coefficient of Pauli string of $A$ \\
        $d_l$ 
        & Coefficient of Pauli string of $A^\dagger A$ \\
        $F$ or $F^{(i)}$ 
        & Forward transform of forcing vector $f$ or $f^{(i)}$ \\
        $D$
        & Size of data instances \\
        $V(\cdot)$ 
        & Parameterized quantum circuit 
        \\ $\mathcal{L}$
        & Cost function\\
        $\theta$ 
        & Parameters for quantum circuit (output of angle network) \\
        $w$ 
        & Angle network parameters \\
        $w^\star$ 
        & Optimal parameters for angle network \\
        $g$ 
        & Angle network \\
        $\gamma$ 
        & Numerator in our loss $\mathcal{L}_{\mathrm{NVQLS}}$ \\
        $\beta$ 
        & Radicand in the denominator of $\mathcal{L}_{\mathrm{NVQLS}}$ \\    
        $l_V$ 
        & Number of layers of the parameterized quantum circuit $V(\cdot)$ \\
          \bottomrule
        \end{tabular}
\end{table}

\section{Derivation of Weak Formulation}
\label{sec:weak_form}

In this section, we present the weak formulation in detail. For a comprehensive overview of the spectral methods employed here, we refer the reader to the standard reference~\cite{shen2011spectral}. We begin by examining second-order elliptic partial differential equations on a bounded domain $\Omega \subset \mathbb{R}^n$ subject to boundary conditions, given a PDE parameter $\epsilon>0$ and an external forcing function $f$:
\begin{equation}
    \begin{aligned}
    -\epsilon \Delta u + \mathcal{F}(u, \nabla u) &=f,
    \quad  x\in \Omega \subset \mathbb{R}^n, \\
    \mathcal{B}(u, \nabla u) &= 0, \quad\quad \quad  x\in \partial \Omega,
    \end{aligned}
    \label{eq:generalpde}
\end{equation}
where $\mathcal{F}$ denotes a possibly nonlinear differential operator and $\mathcal{B}$ represents a boundary operator. For clarity, we first consider the one-dimensional reaction--diffusion case where $\mathcal{F}(u) = u$ and $\Omega = (-1,1)$. For $k = 0, 1, \dots, N-1$, the Legendre--Galerkin weak formulation is given by
\begin{equation}
    \int_\Omega -\epsilon \Delta u(x)\, \phi_k(x)\, dx
    + \int_\Omega u(x)\, \phi_k(x)\, dx
    = \int_\Omega f(x)\, \phi_k(x)\, dx,
\end{equation}
where the basis functions are compact combinations of Legendre polynomials $\{L_k\}$:
\begin{equation}
    \phi_k = L_k + a_k L_{k+1} + b_k L_{k+2}.
\label{eq:compact_combination}
\end{equation}
Here, the exact boundary conditions, including Dirichlet, Neumann, and mixed types, are mathematically enforced by selecting the coefficients $a_k$ and $b_k$. By representing the solution $u$ as a finite linear combination of the basis functions $\{\phi_k\}_{k=0}^{N-1}$, the weak formulation yields
\begin{equation}
    -\epsilon
    \sum_{k=0}^{N-1} \alpha_k
    \left(
    \int_\Omega 
    \phi_j''(x)
    \phi_k(x)
    \, dx \right)
    + 
    \sum_{k=0}^{N-1}
    \alpha_k 
    \left( \int_\Omega
    \phi_j(x)
    \phi_k(x)
    \, dx \right)
    = \int_\Omega f(x)\, \phi_j(x)\, dx,
\end{equation}
for $j=0,1,\dots,N-1$. This produces a linear system $(-\epsilon S+M) \alpha = F$ corresponding to the one-dimensional reaction--diffusion equation, where $S$ denotes the stiffness matrix and $M$ represents the mass matrix. Their entries are specified as follows, for $k,j=0,\dots,N-1$:
\begin{equation}
    S_{kj} = \int_\Omega \phi_j''(x) \phi_k(x) \, dx
    ,\quad
    M_{kj} = \int_\Omega \phi_j(x) \phi_k(x) \, dx,
\end{equation}
with the forward transformed vector given by
\begin{equation}
    F = 
    \begin{bmatrix}
        \tilde{f}_0 & \tilde{f}_1 & \cdots & \tilde{f}_{N-1}
    \end{bmatrix}^T
    ,\quad
    \tilde{f}_k = \int_\Omega f(x) \phi_k(x) \, dx.
\end{equation}
By solving this linear system, one obtains the spectral coefficients
\begin{equation}
    \alpha =
    \begin{bmatrix}
        \alpha_0 & \alpha_1 & \cdots & \alpha_{N-1}
    \end{bmatrix}^T.
\end{equation}
The stiffness matrix $S$ is diagonal and the mass matrix $M$ is symmetric pentadiagonal, whose entries are given by
\begin{equation}
    S_{kj}
    := \int_I \phi_j'' \phi_k w \, dx
    =
    \begin{cases}
    (4k + 6) b_k,
    & j=k, \\
    0,
    & \text{otherwise,}
    \end{cases}
\end{equation}
and
\begin{equation}
    M_{jk} = M_{kj}
    := \int_I \phi_j \phi_k w \, dx
    =
    \begin{cases} 
    \frac{2}{2k + 1} + a_k^2 \frac{2}{2k + 3} + b_k^2 \frac{2}{2k + 5}, & j = k, \\
    a_k \frac{2}{2k + 3} + a_{k+1} b_k \frac{2}{2k + 5}, & j = k + 1, \\
    b_k \frac{2}{2k + 5}, & j = k + 2, \\
    0, & \text{otherwise.}
    \end{cases}
    \label{eq:mass}
\end{equation}
Similarly, for the one-dimensional convection--diffusion equation with Dirichlet boundary conditions, we define the convection matrix $R$ to represent the first spatial derivative $u_x$:
\begin{equation}
\label{eq:R_matrix}
    R_{kj}
    = -R_{jk}
    = \int_I \phi_j' \phi_k w \, dx 
    =
    \begin{cases}
        2 & k=j+1, \\
        -2 & k=j-1, \\
        0 & \text{otherwise.}
    \end{cases}
\end{equation}
For the two-dimensional settings, we define the two-dimensional basis as the tensor product of one-dimensional basis functions:
\begin{equation}
\label{eq:basis_2d}
    \{\phi_k(x)\phi_j(y) \,:\, k,j=0,\dots,N-1\},
\end{equation}
where $\phi_\cdot(\cdot)$ denotes the one-dimensional basis functions, each formed as a compact combination of Legendre polynomials. Consequently, the predicted solution $\hat{u}(x,y)$ on the two-dimensional domain is expressed as a linear combination of these tensor-product basis functions, with coefficients $\hat{\alpha}_{kj}$:
\begin{equation}
    \hat{u}(x,y) =
    \sum_{k=0}^{N-1}
    \sum_{j=0}^{N-1}
    \hat{\alpha}_{kj}
    \phi_k(x)\phi_j(y).
\end{equation}
The weak formulation in two dimensions can be expressed as
\begin{equation}
\label{eq:weak_form_2d}
    -\epsilon
    \iint_\Omega
    \Delta u(x,y) \,
    \phi_k(x) \phi_j(y) \, dx \, dy
    + \iint_\Omega u(x,y) \,
    \phi_k(x) \phi_j(y) \, dx \, dy
    = \iint_\Omega f(x,y) \, 
    \phi_k(x) \phi_j(y) \, dx \, dy,
\end{equation}
or equivalently,
\begin{equation}
    A \alpha = F
    ,\quad \text{where} \quad A = 
    -\epsilon (S \otimes M + M \otimes S) + M \otimes M,
\end{equation}
where $\alpha$ and $F$ are vectors formed by
\begin{equation}
    F=
    (
    \tilde{f}_{0,0}
    ,\tilde{f}_{1,0}
    ,\dots
    ,\tilde{f}_{N-1,0};
    \tilde{f}_{0,1}
    ,\dots
    ,\tilde{f}_{N-1,1};
    \tilde{f}_{0,N-1}
    ,\dots
    ,\tilde{f}_{N-1,N-1}
    )^T,
\end{equation}
and $\otimes$ represents the operation
$
A \otimes B
= (A b_{ij})_{i,j=0,1,\dots,N-1}
$ (i.e., the Kronecker product).

\begin{figure}[t]
    \begin{center}
    \includegraphics[width=\linewidth]{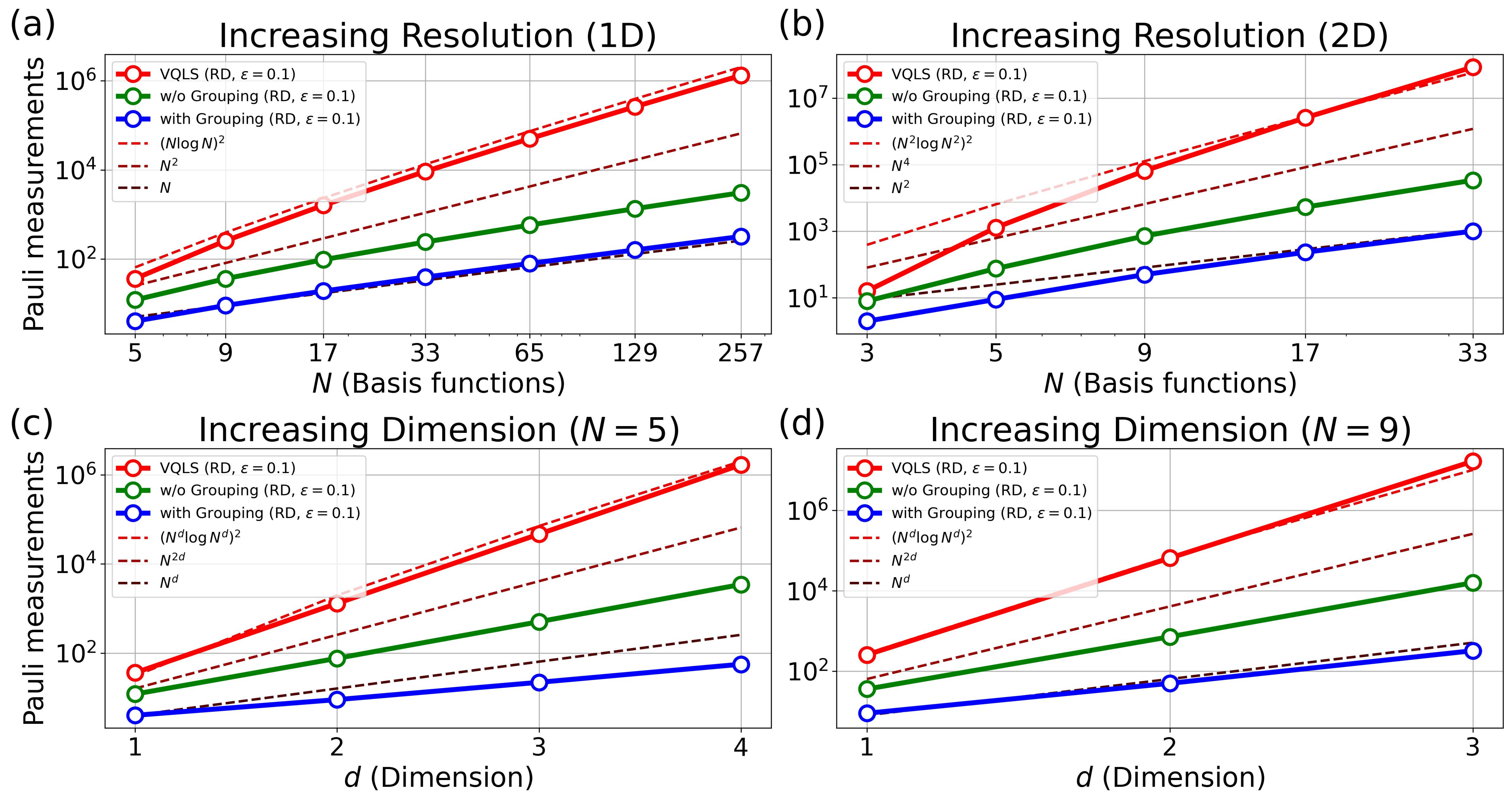}
    \end{center}
    \caption{
    Empirical analysis of the required Pauli measurements. Comparison between NVQLS with measurement grouping (blue), without grouping (green), and the classical VQLS baseline (red). Scaling with respect to (a)-(b) the number of basis functions $N$ for 1D and 2D cases, and (c)-(d) the dimension $d$ for $N=5$ and $N=9$.} \label{fig:figure_advantage_NVQLS_VQLS_RD}
\end{figure}

\begin{figure}[t]
\begin{center}
\includegraphics[width=\linewidth]{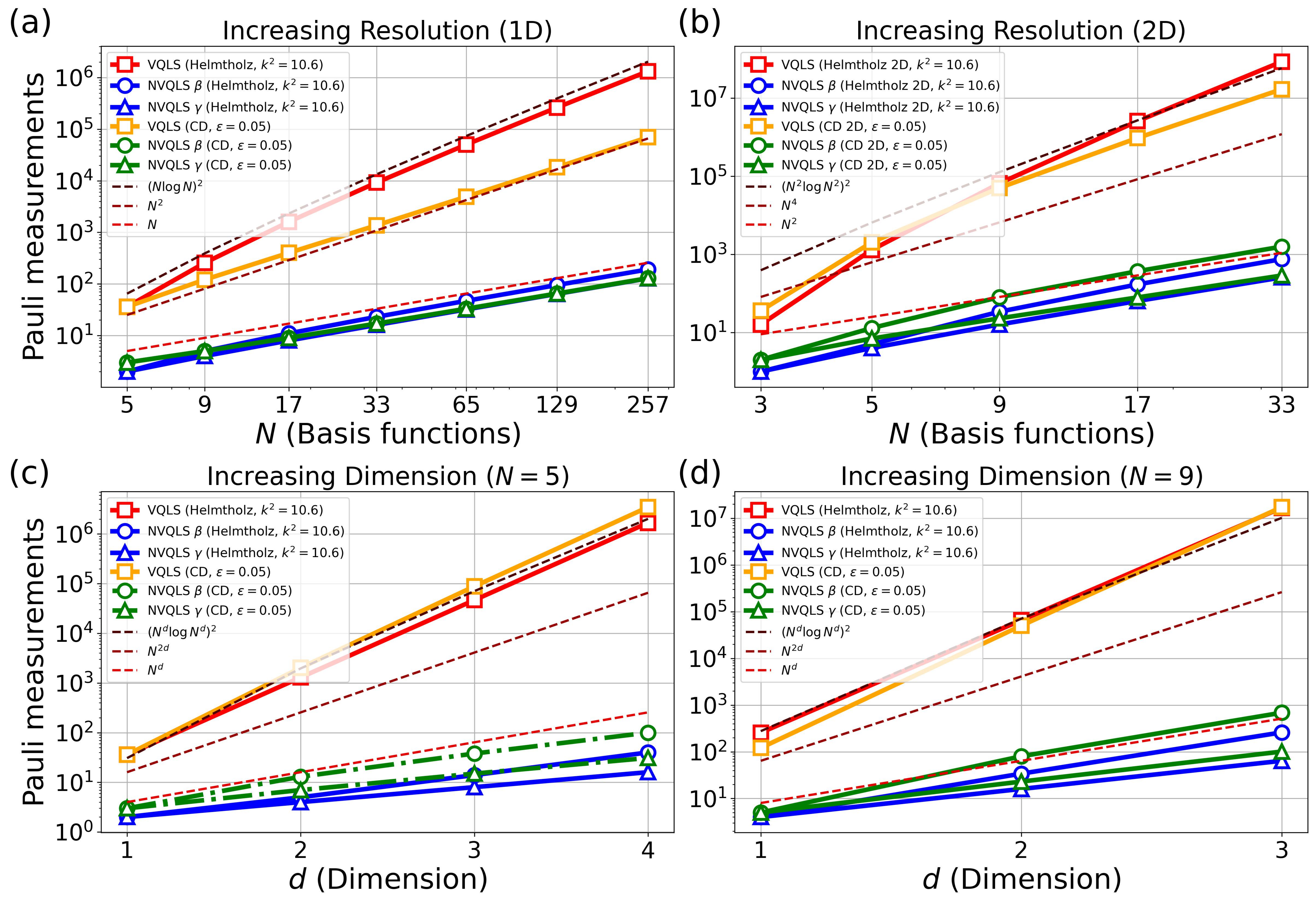}
\end{center}
    \caption{Empirical analysis of the number of required Pauli measurements in NVQLS framework with grouping of commuting measurement operators compared with VQLS of the Helmholtz equation $(k^2 = 10.6)$  and convection diffusion equation $(\epsilon = 0.05)$.
    Top: number of Pauli terms of $A$ for increasing resolution (as a function of the number of basis functions $N$):
    (a) one-dimensional case,
    (b) two-dimensional case.
    Bottom: number of Pauli terms of $A$ for increasing dimension of PDEs  (as a function of the dimension $d$ of PDEs):
    (c) $N=5$,
    (d) $N=9$.
    }
\label{fig:advantage_comparison_CD_Helmholtz}
\end{figure}

\section{Detailed Complexity Analysis}
\label{sec:detailed_complexity_analysis}
We provide a detailed complexity analysis for the results discussed in Sec.~\ref{sec:complexity_analysis}.

\paragraph{Proofs.}
\begin{proof}[Proof of Proposition 3.1]

We analyze the runtime complexity per training iteration of the hybrid NVQLS model. 
Each iteration can be decomposed into a quantum component and a classical component:
$
    T_{\mathrm{per\text{-}iter}}
=
T_{\mathrm{quantum}} + T_{\mathrm{classical}}.
$
This decomposition follows from the hybrid structure of NVQLS: the quantum circuit parameters $\theta$ are generated by the classical neural network with trainable parameters $w$, and gradients are propagated according to
$
{\partial L(\theta(w))}/{\partial w}
=
(\partial L(\theta)/\partial \theta)
(\partial \theta(w)/\partial w).
$
Thus, each training iteration consists of a quantum component, which estimates the loss and its gradients with respect to the quantum parameters, and a classical component, which backpropagates these gradients through the neural network and updates the parameters $w$.

The quantum component can be written as
$
T_{\mathrm{quantum}}
=
T_{\mathrm{prep}}
N_{\mathrm{circuits}}
N_{\mathrm{shots}},
$
where $T_{\mathrm{prep}}$ is the cost of executing a single circuit, $N_{\mathrm{circuits}}$ is the number of circuit evaluations required per iteration, and $N_{\mathrm{shots}}$ is the number of shots required to estimate each observable to precision $\epsilon$. Since $T_{\mathrm{iter}}$ is highly problem- and optimizer-dependent, we focus on the dominant per-iteration cost.

We first analyze the quantum optimization component. The single-circuit execution cost $T_{\mathrm{prep}}$ is proportional to the circuit depth, including both the state-preparation or encoding circuit and the variational ans\"atz. In NVQLS, the dominant contribution comes from the numerator circuit, which contains both the $f$-embedding circuit and the variational ans\"atz. We use logarithmic-depth constructions for both components, yielding
$
T_{\mathrm{prep}} = \mathcal{O}(\log N^d).
$
Although the worst-case depth of the $f$-embedding circuit can scale as $\mathcal{O}(N^d)$, our experiments show that the logarithmic-depth construction is sufficient to represent challenging PDE instances with competitive accuracy; see Appendix~\ref{sec:Hardware-Efficient Training}.

For $N_{\mathrm{circuits}}$, we count the circuit evaluations needed for gradient estimation. A single evaluation of the NVQLS loss involves independent measurements of its numerator and denominator components. This measurement overhead is significantly mitigated by the NVQLS circuit architecture, which facilitates efficient Pauli grouping.
Figure~\ref{fig:figure_advantage_NVQLS_VQLS_RD} and \ref{fig:advantage_comparison_CD_Helmholtz} illustrate the empirical scaling of these numerator and denominator terms, which require $\mathcal{O}(N^d)$ and $\mathcal{O}(N^d\log N^d)$ evaluations, respectively. 
Therefore, the loss-evaluation cost is dominated by the denominator terms and scales as
$
\mathcal{O}(N^d\log N^d).
$
Since the Pauli grouping is structurally impossible in the conventional VQLS loss-evaluation circuit, its observed circuit cost follows $\mathcal{O}((N^d\log N^d)^2)$ scaling.

For gradient estimation, the parameter-shift rule multiplies this base cost by the number of trainable parameters, which scales as $\mathcal{O}((\log N^d)^2)$, yielding
$
N_{\mathrm{circuits}} = \mathcal{O}(N^d(\log N^d)^3).
$
In contrast, SPSA introduces only a constant multiplicative factor, yielding
$
N_{\mathrm{circuits}} = \mathcal{O}(N^d\log N^d).
$
Combining the parameter-shift estimate with $T_{\mathrm{prep}} = \mathcal{O}(\log N^d)$ and $N_{\mathrm{shots}} = \mathcal{O}(1/\epsilon^2)$ gives the following per-iteration quantum optimization cost for NVQLS:
$
T_{\mathrm{quantum}}
=
\mathcal{O}\!\left(
{N^d(\log N^d)^4}/{\epsilon^2}
\right).
$

We now analyze the classical optimization component. In NVQLS, the classical cost arises from backpropagation through the neural network after the quantum gradients have been estimated. The network maps $\mathcal{O}(N^d)$-dimensional input features to an $\mathcal{O}((\log N^d)^2)$-dimensional output vector, corresponding to the quantum parameters. Therefore, the classical optimization cost scales as
$
T_{\mathrm{classical}}
=
\mathcal{O}\!\left(
N^d(\log N^d)^2
\right).
$
Since the quantum optimization term dominates the classical term under the parameter-shift gradient-estimation scheme, the overall per-iteration training cost of NVQLS scales as
$
\mathcal{O}\!\left(
{N^d(\log N^d)^4}/{\epsilon^2}
\right).
$

The runtime can be further reduced by approximating the original linear-system operator with a truncated Pauli expansion, as discussed in Appendix~\ref{sec:truncation method}. Under a target truncation error of order $10^{-3}$, the number of Pauli terms defining the truncated operator is reduced to $\mathcal{O}(N^d)$, compared with $\mathcal{O}(N^d\log N^d)$ for the original Pauli expansion. Applying Pauli grouping to this truncated operator further reduces the total number of circuits required for loss evaluation from $\mathcal{O}(N^d)$ to $\mathcal{O}(\log N^d)$. Under the same assumptions as above, the quantum optimization cost of NVQLS with the truncated Pauli approximation is therefore upper bounded by
$\mathcal{O}\!({(\log N^d)^4}/{\epsilon^2}).$
The classical optimization cost remains unchanged from the non-truncated NVQLS case:
$
\mathcal{O}\!\left(
N^d(\log N^d)^2
\right).
$
Therefore, when the truncated Pauli approximation is used, the classical optimization term dominates, scaling as $\mathcal{O}\!({(\log N^d)^4}/{\epsilon^2})$.
\end{proof}

\begin{proof}[Proof of Proposition 3.2]
We analyze the per-iteration classical memory cost of hybrid NVQLS when the
variational quantum circuit is executed on quantum hardware.
In this setting, the quantum
state vector is not stored classically. The circuit parameters are generated by
the classical neural network as $\theta=\theta(w)$, and gradients with respect
to the trainable classical parameters are computed by the chain rule,
$\partial L(\theta(w))/\partial w
=
(\partial L(\theta)/\partial \theta)
(\partial \theta(w)/\partial w)$.
Thus, the classical memory associated with the quantum component only needs to
store the generated circuit parameters $\theta(w)$ and, if gradients are stored
explicitly, the corresponding quantum gradient vector $\partial L/\partial
\theta$.

Let $M_{\rm total}
=
\mathcal{O}(M_{\rm q}+M_{\rm c}+M_{\rm grad}+M_{\rm act})$, where $M_{\rm q}$ denotes the
temporary classical memory required to store $\theta(w)$ and $\partial
L/\partial \theta$, $M_{\rm c}$ denotes the memory for the classical
neural-network parameters $w$, $M_{\rm grad}$ denotes the memory for gradients
and optimizer states, and $M_{\rm act}$ denotes the activation memory required
for backpropagation through the classical network.

Under the assumptions of Proposition~3.1, the ans\"atz uses $n=\mathcal{O}(\log N^d)$ qubits
and has depth $D=\mathcal{O}(\log N^d)$. Assuming that each layer contains $\mathcal{O}(n)$
parametrized gates, the number of generated circuit parameters is
$P=\mathcal{O}(nD)=\mathcal{O}((\log N^d)^2)$. With constant-precision storage per scalar, the
corresponding temporary quantum memory is therefore
$M_{\rm q}=\mathcal{O}(P)=\mathcal{O}((\log N^d)^2)$.

The classical neural network takes an input feature vector of dimension
$\mathcal{O}(N^d)$ and outputs $P=\mathcal{O}((\log N^d)^2)$ circuit parameters. For a
constant-depth fully connected network whose hidden-layer widths are at most
$\mathcal{O}((\log N^d)^2)$, the number of trainable classical parameters is dominated by
the first linear layer and satisfies
$M_{\rm c}=\mathcal{O}\!\left(N^d(\log N^d)^2\right)$.

For constant batch size and standard first-order optimizers with a constant
number of auxiliary states per trainable parameter, the memory for gradients and
optimizer states satisfies $M_{\rm grad}=\mathcal{O}(M_{\rm c}+M_{\rm q})$. Moreover,
since the classical network has constant depth and hidden-layer widths at most
$\mathcal{O}((\log N^d)^2)$, its activation memory during backpropagation is
$M_{\rm act}=\mathcal{O}\!\left(N^d+(\log N^d)^2\right)=\mathcal{O}(M_{\rm c})$.

Combining these estimates gives
$M_{\rm total}
=
\mathcal{O}(M_{\rm q}+M_{\rm c}+M_{\rm grad}+M_{\rm act})
=
\mathcal{O}(M_{\rm q}+M_{\rm c})$.
Since $M_{\rm q}=\mathcal{O}((\log N^d)^2)$ is dominated by
$M_{\rm c}=\mathcal{O}\!\left(N^d(\log N^d)^2\right)$, we obtain
$M_{\rm total}
=
\mathcal{O}\!\left(N^d(\log N^d)^2\right)$.
\end{proof}

\paragraph{Comparison with classical neural operator methods.}
For the classical unsupervised neural operator methods considered in this work, namely SCLON and PI-DON, the per-iteration training cost is dominated by neural-network forward and backward passes. In SCLON, dense fully connected layers map input features of dimension proportional to $N^d$ to a spectral coefficient vector of dimension $N^d$, leading to $\mathcal{O}(N^{2d})$ arithmetic operations per iteration.
In contrast, PI-DON evaluates the residual by pointwise sampling rather than weak-form integration. Nevertheless, assuming the same number of shared nodal points, the dominant matrix multiplication still requires $\mathcal{O}(N^{2d})$ arithmetic operations per iteration.

The memory costs of SCLON and PI-DON are also dominated by the dense neural-network layers. Assuming constant batch size, constant network depth, and constant-precision storage, the memory required for parameters, gradients, and optimizer states is proportional to the number of trainable weights. In SCLON, the dominant fully connected mapping from $\mathcal{O}(N^d)$ input features to an $\mathcal{O}(N^d)$ spectral coefficient vector requires $\mathcal{O}(N^d)\times \mathcal{O}(N^d)=\mathcal{O}(N^{2d})$ trainable parameters. The activation memory is only $\mathcal{O}(N^d)$ per layer and is therefore dominated by the parameter memory under the constant-depth assumption. Hence, the per-iteration memory cost of SCLON is $\mathcal{O}(N^{2d})$. Similarly, for PI-DON, pointwise residual evaluation changes the form of the loss evaluation but not the dominant dense neural-network storage cost under the same number of shared nodal points. The dominant matrix multiplication corresponds to storing $\mathcal{O}(N^{2d})$ trainable weights, and the associated gradients and optimizer states have the same asymptotic order. Therefore, the per-iteration memory cost of PI-DON is also $\mathcal{O}(N^{2d})$.



\section{Derivation of Phase-Aware Loss Function}
\label{sec:loss_derivation}

To solve the linear system $A\alpha = F$, our variational quantum framework prepares a parameterized state $\ket{\hat{\alpha}(w)}$. The action of the operator $A$ on this state yields an unnormalized state $A\ket{\hat{\alpha}(w)}$. We define the corresponding normalized solution state as:$$\ket{\psi} = \frac{A\ket{\hat{\alpha}(w)}}{\sqrt{\bra{\hat{\alpha}(w)}A^\dagger A\ket{\hat{\alpha}(w)}}}$$The primary objective is to find the optimal parameters $w$ such that the prepared state $\ket{\psi}$ exactly aligns with the normalized target state $\ket{F}$, which implies $\ket{\psi} = \ket{F}$. In terms of their inner product (overlap), this target alignment is achieved when:$$\braket{F|\psi} = 1$$Traditional overlap-based loss functions often rely on maximizing the squared magnitude $\left| \braket{F|\psi} \right|^2$.
However, this introduces a critical phase ambiguity, as any state $\ket{\psi} = e^{i\theta}\ket{F}$ would still minimize the loss, leading to incorrect solution representations. To resolve this phase ambiguity without introducing the significant measurement overhead associated with computing complex phases, we propose a phase-aware loss function based solely on the real part of the overlap. Since both $\ket{\psi}$ and $\ket{F}$ are normalized states, the Cauchy-Schwarz inequality restricts their complex overlap to the unit disk: 
$$
    |\braket{F|\psi}| \le 1
$$
Expressing the overlap in terms of its real and imaginary components, we have:
$$
    |\mathrm{Re}\braket{F|\psi}|^2 + |\mathrm{Im} \braket{F|\psi}|^2 \le 1
$$
From this bounded geometry, it mathematically follows that maximizing the real part strictly constrains the imaginary part. Specifically, if we enforce $\mathrm{Re}\braket{F|\psi} = 1$, the inequality mandates that $(\mathrm{Im}\braket{F|\psi})^2 \le 0$, which is only physically satisfied when $\mathrm{Im}\braket{F|\psi} = 0$.
$$\mathrm{Re}\braket{F|\psi} = 1 \quad \iff \quad \braket{F|\psi} = 1 \quad \text{(with } \mathrm{Im}\braket{F|\psi} = 0\text{)}$$
Therefore, explicitly evaluating the imaginary component is mathematically redundant for reaching the exact global minimum. We can define our loss function by simply penalizing the deviation of the real part of the overlap from 1:
$$
    \mathcal{L}_{\mathrm{NVQLS}}(f;w) = 1 - \text{Re}\braket{F|\psi}
$$
Substituting the definition of $\ket{\psi}$ into the loss function yields:
$$
    \mathcal{L}_{\mathrm{NVQLS}}(f;w) = 1 - \frac{\text{Re}(\bra{F}A\ket{\hat{\alpha}(w)})}{\sqrt{\bra{\hat{\alpha}(w)}A^\dagger A\ket{\hat{\alpha}(w)}}}
$$
Finally, substituting the unitary decompositions of $A$ and $A^\dagger A$, defined as $A = \sum_{l=1}^{L} c_{l} A_{l}$ and $A^\dagger A = \sum_{l=1}^{R} d_{l} A_{l}$, and averaging over a dataset of size $D$ with target states $\ket{F^{(i)}}$, we arrive at the computable form of the cost function in Eq.~\eqref{eq:loss_NVQLS}. We also used an unnormalized cost function $\hat{\mathcal{L}}_{\mathrm{NVQLS}}$ for training stability defined as
\begin{equation*}
\label{eq:loss_NVQLS_hat}
\begin{aligned}
    & \hat{\mathcal{L}}_{\mathrm{NVQLS}}(w)=
    \\&\frac{1}{D}\sum_{i=1}^{D} 
    \left(
    \mathrm{Re}\left(
    \sum_{l=1}^L c_l \langle F^{(i)} | A_l | {\hat{\alpha}(f^{(i)} ;w)} \rangle \right)-
    {\sqrt{ 
    \sum_{l=1}^R d_l
    \langle {\hat{\alpha}(f^{(i)} ;w)} | A_l | {\hat{\alpha}(f^{(i)} ;w)} \rangle}
    }
    \right)^2.
\end{aligned}
\end{equation*}
The rescaled loss is algebraically equivalent to the original cost function because both formulations share the exact same global minimum. By adopting this quadratic form, we eliminate numerical instabilities caused by gradient divergence, leading to significantly more robust and reliable convergence during training. In practice, this loss function often provides better convergence, whereas the fractional form in Eq.~\eqref{eq:loss_NVQLS} can sometimes become unstable during training.

\section{Training Details}
\label{sec:training_details}

In this section, we provide the details of the model architecture used in our experiments.  All simulation training runs are performed using PennyLane~\cite{bergholm2020pennylane}, supported by the JAX framework~\cite{bradbury2018jax}. Our model employed the ans\"atz $V(\theta)$ to prepare $|\alpha\rangle=V(\theta)|0\rangle$ and utilize the classical neural network $g$ to represent $\theta$ as $\theta=g(w;F)$ with its parameters $w$ for given forcing vector $f$. 
Furthermore, two types of loss functions are employed in this work.

\begin{figure}[t]
\begin{center}
    \includegraphics[width=\linewidth]{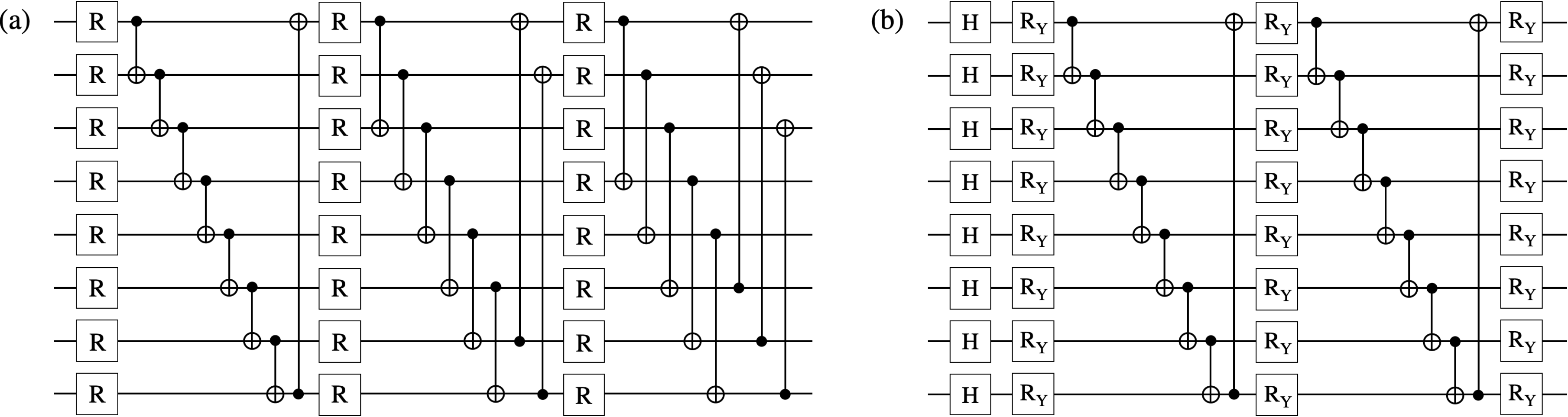}
\end{center}
\caption{Ans\"atze used in our study. (a) Strongly Entangling Layer, (b) Hardware efficient RY ans\"atz.
}
\label{fig:ansatz}
\end{figure}

\subsection{Quantum Circuit Ans\"atz}
\label{sec:ansatz}

We used two types of ans\"atz $V(\theta)$: the strongly entangling circuit and the hardware-efficient RY ans\"atz. 
Figure~\ref{fig:ansatz} describes each ans\"atz structure. The strongly entangling layer consists of rotation gates $R$ and $\mathrm{CNOT}$ gate.
The rotation gate $R$ can represent a general rotation by using three parameters $\phi,\theta,\omega$ through two types of single qubit rotation gate $R_Y$ and $R_Z$, defined as $R\equiv R(\phi,\theta,\omega)=R_Y(\phi)R_Z(\theta)R_Y(\omega)$.
This ans\"atz contains strong entanglements where each qubit is entangled with two other qubits with $\mathrm{CNOT}$ gates in each layer, employing the different patterns of entanglements per layer.
Due to its characteristics, this possess high representation capacity which can be quantified with high expressibility and high entangling capacity.
The number of parameters for this ans\"atz is $(3\times n\times l_V)$ which is proportional to the number of qubits $n$ and the number of the ans\"atz layer $l_V$.

The Hardware efficient RY ans\"atz is composed of Hadamard gates $H$, single qubit rotation gates on y-axis $R_Y$, and $\mathrm{CNOT}$ gates.
The Hadamard gates are applied first on each qubit, and rotation gates $R_Y\equiv R_Y(\theta)$ are parameterized with $\theta$. In each layer, the nearest neighbour qubits are entangled by $\mathrm{CNOT}$ as well as the first and the last qubits.
This ans\"atz is hardware efficient. Moreover, it maps the initial state $|0\rangle$ to real-valued vector $|\alpha\rangle\in \mathbb{R}^{2^n}$ so we don't need to consider the imaginary part when extracting the vector components from $|\alpha\rangle$.
The number of parameters for this ans\"atz is $(n\times l_V)$, also proportional to both $n$ and $l_V$.

\begin{table}[t]
    \centering
    \caption{Details of the quantum circuit ans\"atz structure for each PDE benchmark in the main experiment. Here, $n$ denotes the number of qubits and $l_V$ represents the number of layers in the ans\"atz.}
    \label{tab:ansatz_details_refined}
    \begin{tabular}{llcc}
        \toprule
        \textbf{PDE} & \textbf{Ans\"atz Structure} & \textbf{$n$} & $l_V$ \\
        \midrule
        1D RD              & StronglyEntanglingLayers & 5 & 12 \\
        1D Helmholtz       & StronglyEntanglingLayers & 5 & 12 \\
        1D CD              & StronglyEntanglingLayers & 5 & 12 \\
        1D Wave            & StronglyEntanglingLayers & 4 & 12 \\ 
        \midrule
        2D RD              & StronglyEntanglingLayers & 6 & 12 \\
        2D Helmholtz       & StronglyEntanglingLayers & 6 & 12 \\
        2D CD              & StronglyEntanglingLayers & 6 & 20 \\
        \midrule
        Joint Helmholtz    & StronglyEntanglingLayers & 6 & 15 \\
        \bottomrule
    \end{tabular}
\end{table}

\subsection{Classical Angle Network}

As the angle networks for one-dimensional PDEs, we employ feed-forward (FF) neural networks with Rectified Linear Unit (ReLU)~\cite{agarap2019deeplearningusingrectified} activation functions.
To ensure stable and fast convergence, the L-BFGS optimizer~\cite{Liu1989OnTL} is utilized for the cost function.
For the operator learning with joint inputs, a deeper and wider network structure with six FF layers was utilized to enhance the generalization ability of the angle networks.
For two-dimensional PDEs, we use a hybrid network combining three convolutional neural network (CNN) layers followed by three FF layers, with the Gaussian error linear units (GELU)~\cite{hendrycks2016gaussian} activation functions.
For the convection--diffusion equation with preconditioner, the adaptive moment estimation (Adam)~\cite{kingma2017adam} optimizer is utilized. For hyper-parameter optimization, we adopted a heuristic, progressive approach, starting from a minimal architecture and increasing width or depth only when it improved generalization on validation PDE instances. Due to the shallow quantum circuits in NVQLS, small architectures sufficed for stable training.

\begin{table}[t]
    \caption{Architectural details of the angle network for the primary PDE benchmarks. For 1D cases, the network consists of fully-connected (FC) layers, whereas a hybrid architecture combining convolutional (CNN) and FC layers is employed for 2D problems.}
    \label{tab:angle_net_details}
  \begin{center}
        \begin{tabular}{lcccccr}
          \toprule
        PDE
        & NN
        & LAYER
        & $\#$ of params
        & ACTIVATION
        \\ \midrule
        1D RD
        & FC
        & 4
        & 96,732 
        & \makecell[l]{GeLU}
    \\
        1D Helmholtz
        & FC
        & 4
        & 92,052
        & \makecell[l]{GeLU}
    \\
        1D CD
        & FC
        & 4
        & 92,052
        & \makecell[l]{GeLU}
        \\
        1D Wave
        & CNN+FC
        & 3+4
        & 55,612 
        & \makecell[l]{ReLU}
    \\
        \midrule
        RD 2D
        & CNN+FC
        & 3+3
        &  182,886 
        & \makecell[l]{ReLU}
    \\
        Helmholtz 2D
        & CNN+FC
        & 3+3
        &  89,226 
        & \makecell[l]{ReLU}
    \\
        CD 2D
        & CNN+FC
        & 3+3
        & 84,582
        & \makecell[l]{ReLU}
    \\
        \midrule
        Joint Helmholtz
        & CNN+FC
        & 3+3
        & 89,482
        & \makecell[l]{ReLU}
    \\
          \bottomrule
        \end{tabular}
  \end{center}
\end{table}

\begin{table}[ht]
\centering
\caption{Specifications of GPU and CPU hardware used for computation}
\label{tab:hardware_specs}
\begin{tabular}{ll}
\toprule
\textbf{Specification} & \textbf{Value} \\ 
\midrule
CPU Model Name & Intel(R) Xeon(R) Gold 6348 CPU @ 2.60GHz \\
CPU(s) & 56 \\
Thread(s) per core & 2 \\
Core(s) per socket & 28 \\
Socket(s) & 1 \\
NUMA node(s) & 1 \\
CPU MHz (Max) & 3500.00 \\
L1d \& L1i cache & 1.3 MiB / 896 KiB \\
L2 cache & 35 MiB \\
L3 cache & 42 MiB \\
RAM & 251 GiB \\
\midrule
GPU Model name & NVIDIA GeForce RTX 3090 \\
CUDA version & 12.4 \\
GPU(s) & 4  \\
GPU Architecture & NVIDIA Ampere \\
Dedicated Memory Size (per GPU) & 24 GB \\
Peak FP32 Performance* & 35.6 TFLOPs \\
Peak Memory Bandwidth* & 936 GB/s \\
\bottomrule
\end{tabular}
\begin{flushleft}
\footnotesize $^*$ Values based on the official NVIDIA specification sheet for the RTX 3090.
\end{flushleft}
\end{table}


\begin{table}[t]
\centering
\caption{Generation details for the input PDE instances for benchmarks in the main experiment.}
\label{tab:instance_generation}
\begin{tabular}{llccc}
    \toprule
    \textbf{PDE} & \textbf{Type} & \textbf{$N$ ($n$)} & \textbf{\makecell[c]{Forcing \\ Distribution}} & \textbf{\makecell[c]{PDE Coefficient \\ Distribution}} \\
    \midrule
    1D RD             & Shallow & 32 (5) & $\theta \sim U[0, 2\pi)$ & $0.1$ \\
    1D Helmholtz      & Shallow & 32 (5) & $\theta \sim U[0, 2\pi)$ & $4.7$ \\
    1D CD             & Shallow & 32 (5) & $\theta \sim U[0, 2\pi)$ & $0.1$ \\
    1D Wave           & Specified in Eq. \eqref{eq:wave_forcing} & 16 (4) & $w \sim U[1, 2)$ & $-$ \\
    \midrule
    2D RD             & General & 8 (6)  & $h_i,m_i \sim U[0, 1)$  & $0.1$ \\
    2D Helmholtz      & General & 8 (6)  & $h_i,m_i \sim U[0, 1)$  & $8.9$ \\
    2D CD             & General & 8 (6)  & $h_i,m_i \sim U[0, 1)$  & $0.1$ \\
    \midrule
    Joint Helmholtz   & General & 8 (6)  & $h_i,m_i \sim U[0, 1)$  & $U(4, 4.05)$ \\
    \bottomrule
\end{tabular}
\end{table}

\section{Performance Metrics}
\label{sec:performance_metric}

We used numerical solutions based on a spectral method as the ground truth, and both were evaluated on the same grid when computing the errors. Classical spectral convergence results~\cite{shen2011spectral, doi:10.1137/1.9780898719598, Burden1989} ensure that the approximation error decays extremely fast so that, with a suitable number of basis functions, the numerical error becomes effectively close to the limits of machine precision, typically around $10^{-12} \sim 10^{-14}$.

In our experiments, for the qubit numbers ($n = 4, 5$), the corresponding spectral basis sizes ($N = 17, 33$) are sufficient to reach this accuracy regime, and the numerical solution used for error computation therefore achieves high-accuracy. Any interpolation of the true solution is used only for visualization in the plots and does not affect the underlying error calculations.

To measure the prediction errors, we employ three metrics: batch-wise mean absolute error (MAE), batch-wise relative $L_2$ error, and batch-wise relative $L_\infty$ error. Given a collection $\{\hat{u}^{(i)}\}_{i=1}^D$ of predicted solutions and the corresponding true collection $\{u^{(i)}\}_{i=1}^D$, the batch-wise mean absolute error (MAE) is defined by
\begin{equation}
\label{eq:mae}
    \mathrm{MAE}
    = 
    \frac{1}{D (N+1)} \sum_{i=1}^{D} \sum_{j=0}^{N} 
    \big| \hat{u}^{(i)}(x_j) - u^{(i)}(x_j) \big|,
\end{equation}
where $D$ is the number of data instances, $N$ corresponds to the number of basis functions (or spatial points), and $\hat{u}^{(i)}(x_j)$ and $u^{(i)}(x_j)$ are the predicted and true solutions at the collocation point $x_j$, respectively. The batch-wise  relative $L_2$ error is given by
\begin{equation}
\mathrm{RelL_2}
=
\frac{1}{D} \sum_{i=1}^D
\frac{ \| \hat{u}^{(i)} - u^{(i)} \|_2 }{ \| u^{(i)} \|_2 }
= 
\frac{1}{D} \sum_{i=1}^D
\frac{ \sqrt{ \int | \hat{u}^{(i)} - u^{(i)} |^2 dx } }{ \sqrt{ \int | u^{(i)} |^2 dx } },
\end{equation}
where the integrals are computed using the Legendre–Gauss–Lobatto (LGL) quadrature. Finally, the batch-wise relative $L_\infty$ error is defined by
\begin{equation}
\mathrm{RelL_\infty}
=
\frac{1}{D} \sum_{i=1}^D
\frac{ \| \hat{u}^{(i)} - u^{(i)} \|_\infty }{ \| u^{(i)} \|_\infty }
= 
\frac{1}{D} \sum_{i=1}^D
\frac{ \max_{0\leq j\leq N} | \hat{u}^{(i)}(x_j) - u^{(i)}(x_j) | }{ \max_{0\leq j\leq N} | u^{(i)}(x_j)  |}.
\end{equation}

\begin{figure}[t]
    \begin{center}
    \includegraphics[width=\linewidth]{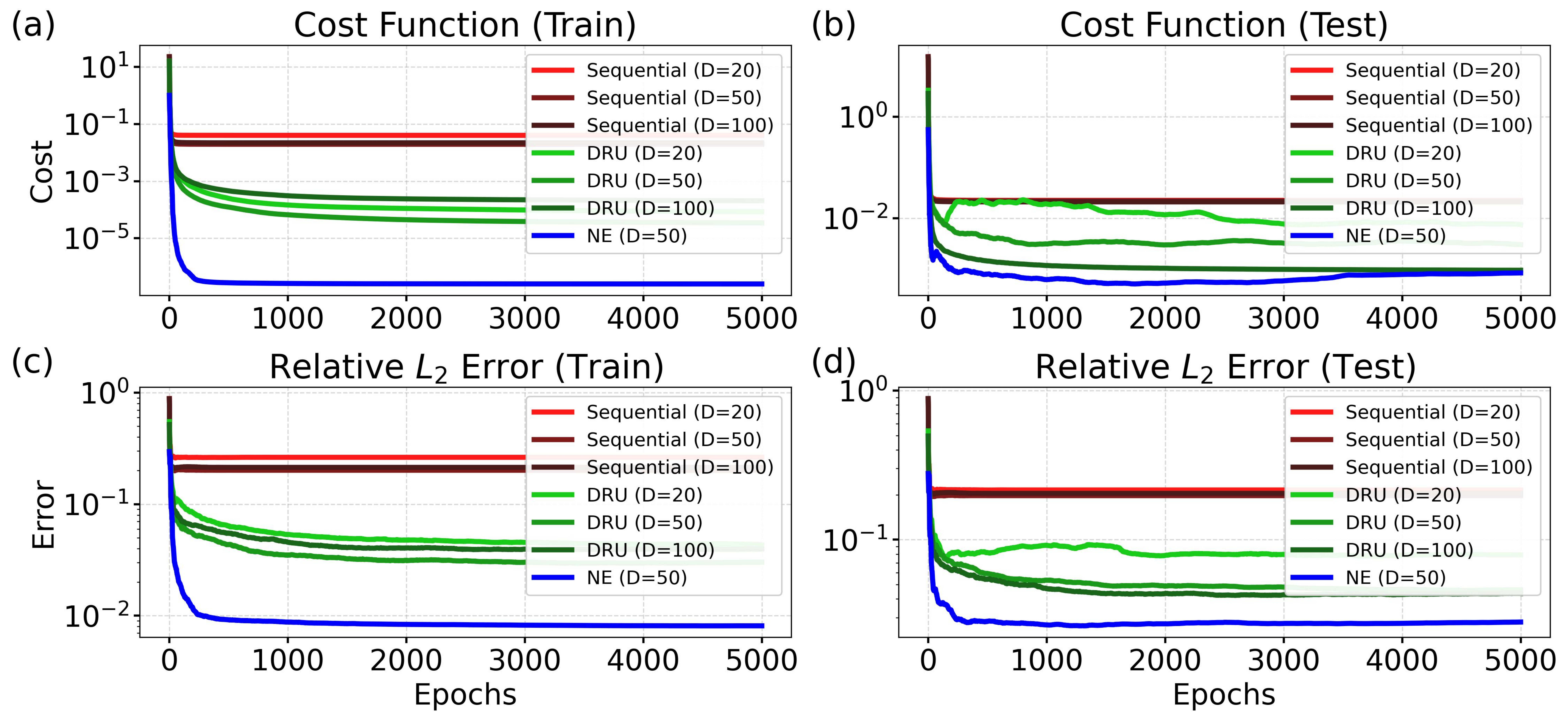}
    \end{center}
    \caption{
    Efficiency analysis of neural embedding under comparable parameter budgets. The plots illustrate histories over training epochs for (a) training cost, (b) testing cost, (c) relative $L^2$ training error, and (d) relative $L^2$ testing error. We compare NVQLS (1,148 parameters, depth 2) against the quantum baselines (1,200 parameters, depth 100). The results demonstrate that neural embedding provides enhanced generalization and performance even with a significantly shallower circuit.
    }
    \label{fig:efficient_ablation}
\end{figure}

\section{Ablation Study of Neural Embedding}
\label{sec:ablation_neural_embedding}

To evaluate the generalization capability of Neural Embedding (NE), we conducted an ablation study comparing our method against baseline approaches. The first baseline is defined as an extension of VQLS with a sequential data embedding scheme, formulated as: 
\begin{equation}
    V(\theta,f) = V(\theta)U(f),
\end{equation}
where the forcing function information is encoded into the quantum circuit using repetitive angle embedding. We also compared our approach with Data Re-uploading (DRU), a latest quantum embedding technique. DRU enhances circuit expressivity by interleaving data encoding $U(f)$ and trainable layers $V(\theta)$, represented as: 
$$
    V(f, \theta) = U(f, \theta) V_{\text{ans\"atz}}(\theta), \quad \text{where} \quad U(f, \theta) = \prod_{i=1}^{l_{\text{DRU}}} [V_i(\theta_i) U_i(f)]
$$
For a rigorous comparison across varying dataset sizes ($D \in \{10, 20, 100\}$), we utilized $l_{\text{DRU}}=50$ re-uploading layers and an additional 50 Strongly Entangling Layers as the variational ans\"atz. Both models were optimized using the L-BFGS optimizer to ensure rapid convergence, minimizing the phase-aware overlap cost function (Eq.~\ref{eq:loss_NVQLS}).

In the first experiment, we evaluated the optimal configuration for each method without constraining the number of trainable parameters. Consequently, NVQLS with NE utilized 7,780 parameters with an ans\"atz depth of 10, while the baseline (without NE) employed 1,200 parameters with a depth of 100. As illustrated in Fig.~\ref{fig:ablation_neural_embedding}, the optimization difficulty for both quantum baselines escalates significantly as the number of instances increases ($D = 10, 20, \text{ and } 100$). In contrast, NVQLS with NE effectively handles $D = 50$ training instances and successfully generalizes across 200 test samples, achieving a substantially lower relative error compared to the baselines. These results underscore that simply extending VQLS without a neural network component is insufficient for effective multi-instance learning.

Furthermore, we evaluated the utility of NE by comparing generalization performance under a comparable parameter setting. In this configuration, NVQLS with NE only utilized 1,148 trainable parameters with a drastically reduced circuit depth of 2, whereas the baseline maintained its previous configuration (1,200 parameters/ 100 depth). Figure~\ref{fig:efficient_ablation} presents the corresponding loss curves and error histories. The results demonstrate that NVQLS with NE outperforms the baseline even with fewer parameters. This underscores the utility of our approach, as it achieves superior performance with significantly reduced model complexity.

\section{Efficient Circuit Implementation}
\label{sec:Hardware-Efficient Training}
This appendix provides empirical support for the state preparation assumption used in the complexity analysis. We do not claim that arbitrary amplitude embeddings are shallow or that state preparation is solved in general. Instead, we test whether the structured forcing functions are compatible with shallow circuit preparation. This would allow the input states used by NVQLS to avoid the generic worst-case cost of quantum state preparation.

\textbf{Shallow Embedding.}
To examine this point, we consider a shallow real-valued state-preparation family based on single-qubit $R_y$ rotations. We define an angle vector $\vtheta=(\theta_1,\dots,\theta_n)$ and prepare the product state
\begin{equation}
\lvert \psi(\boldsymbol{\theta})\rangle
= \bigotimes_{i=1}^n R_y(\theta_i)\lvert 0\rangle
\quad\text{with}\quad
R_y(\theta)=\exp\!\left(-\tfrac{i}{2}\theta\,Y\right),
\end{equation}
where $Y$ denotes the Pauli-$Y$ operator. Let $\va(\boldsymbol{\theta})\in\mathbb{R}^{K}$ denote the real amplitudes of $\lvert \psi(\boldsymbol{\theta})\rangle$ in the computational basis, where $K=2^n$ is the dimension of the quantum state. We set
\begin{equation}
F_k \;=\; a_k(\vtheta) \,,\qquad k=0,\dots, K-1,
\end{equation}
so that $F\in\mathbb{R}^{K}$. The dataset is generated classically by sampling $\vtheta$, computing $F$, and storing the resulting pairs $(\boldsymbol{\theta}, F)$.

\textbf{Data Generation for General Setting.} To evaluate the generalization performance, we generate a diverse set of input forcing functions using random linear combinations of trigonometric functions. For 1D cases, the forcing functions are defined as:
\begin{equation}
\label{eq:forcing_1d}
f^{(i)}(x) = h_1^{(i)} \sin(m_1^{(i)} x) + h_2^{(i)} \cos (m_2^{(i)} x),
\end{equation}
where the parameters $h_j^{(i)}$ and $m_j^{(i)}$ are independently sampled from uniform distributions. These functions, along with their corresponding predicted solutions, are evaluated at nodal points. Similarly, for two-dimensional PDEs, the forcing functions are constructed as:
\begin{equation}
\label{eq:forcing_2d}
\begin{aligned}
f^{(i)}(x, y) = h_1^{(i)} \sin(m_1^{(i)} (x+y)) + h_2^{(i)} \cos (m_2^{(i)} (x+y)),
\quad i=1,2,\dots,D.
\end{aligned}
\end{equation}
Table~\ref{tab:instance_generation} provides further details regarding the data instance generation, including the specific ranges for the parameters and the PDE-specific coefficients such as the diffusion parameter $\epsilon$ and wave number $k$.

\begin{figure}[ht]
\begin{center}
    \includegraphics[width=\linewidth]{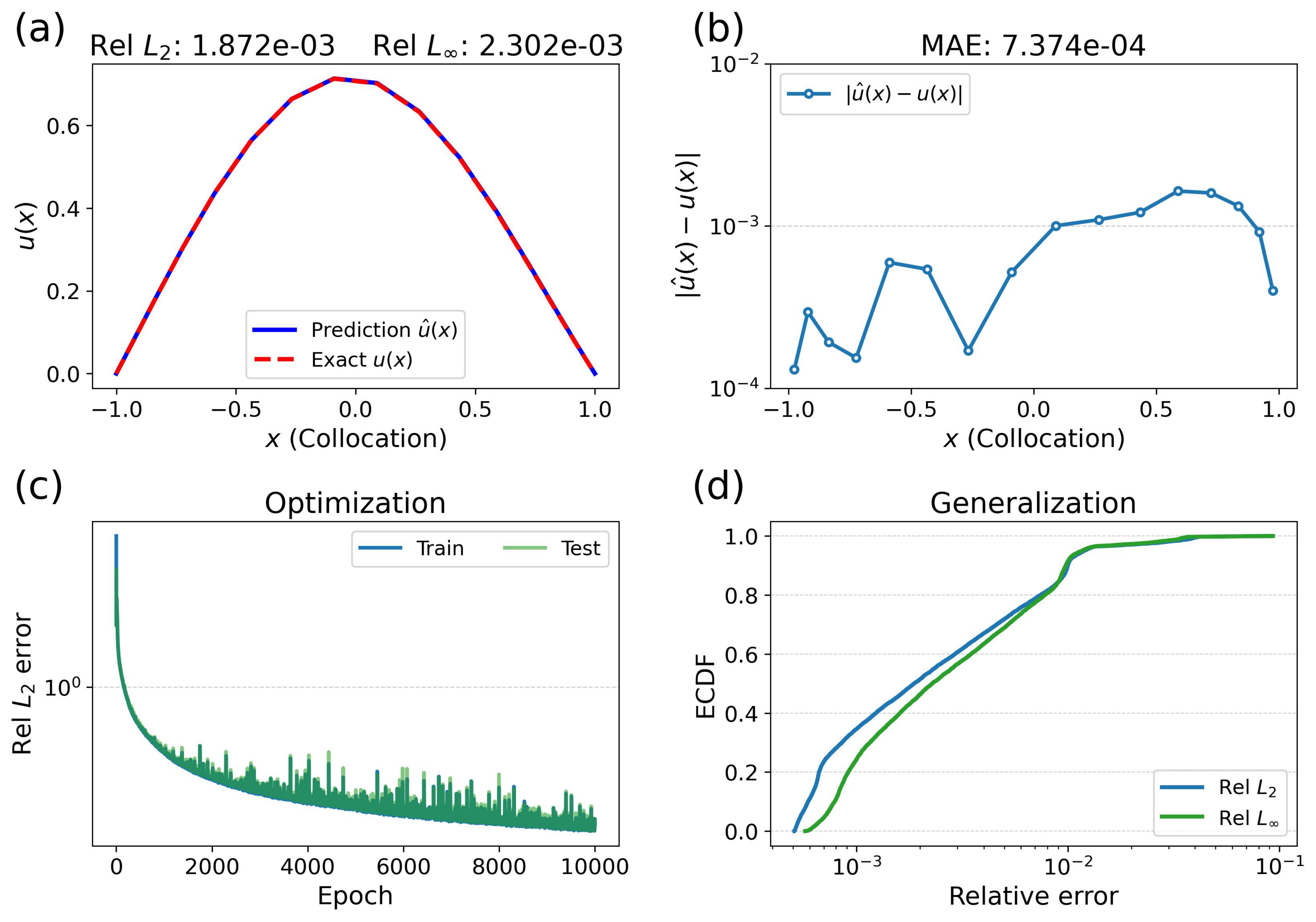}
\end{center}
\caption{
(a) Representative OOD prediction of the solution $u(x)$ for a forcing function in Eq.~\ref{eq:forcing_1d}.
(b) Absolute error $|\hat{u}(x)-u(x)|$ for the same example.
(c) Train and test relative $L_2$ errors on the shallow $R_y$ construction.
(d) Empirical cumulative distribution functions (ECDFs) of relative $L_2$ and $L_\infty$ errors on $10{,}000$ OOD forcing functions.
}
\label{fig:shallow_amplitude_embedding}
\end{figure}

\textbf{OOD Experiment.} We first train NVQLS on transformed forcing vectors generated by the $R_y$ construction. We then evaluate the trained model out of distribution (OOD) on $10{,}000$ forcing functions in Eq.~\ref{eq:forcing_1d}. This OOD evaluation tests whether a model trained using input states prepared by shallow circuits generalizes to the structured trigonometric forcing functions. The experiment is conducted on the one-dimensional Helmholtz equation with Dirichlet boundary conditions using a 4-qubit system corresponding to 18 collocation points, including the two boundary points. Each rotation angle is sampled independently from $U[0,2\pi)$, and the angle network is a fully connected neural network with $21{,}060$ parameters. 
Figure~\ref{fig:shallow_amplitude_embedding} summarizes this evaluation. The model achieves low relative $L_2$ error on the $10{,}000$ OOD forcing functions. This suggests that the structured trigonometric forcing functions used in the main experiments are compatible with shallow circuit preparation, providing empirical support for the state-preparation assumption used in the complexity analysis.

\section{Truncation Method}
\label{sec:truncation method}
To further improve computational efficiency, we approximate the target linear system $A\alpha=F$ by truncating the Pauli decomposition of $A$, which results in a reduced system $\tilde{A}\tilde{\alpha}=F$. Specifically, instead of using the full decomposition $A=\sum^{L}_{l=1} c_l A_l$, we retain only $L' < L$ Pauli terms and define $\tilde{A}=\sum^{L'}_{l=1} c_l A_l$. 
Here, $\tilde{\alpha}$ denotes the solution obtained from the truncated system. This truncation significantly reduces the computational cost of the loss evaluation.

\begin{figure}[t]
\begin{center}
    \includegraphics[width=\linewidth]{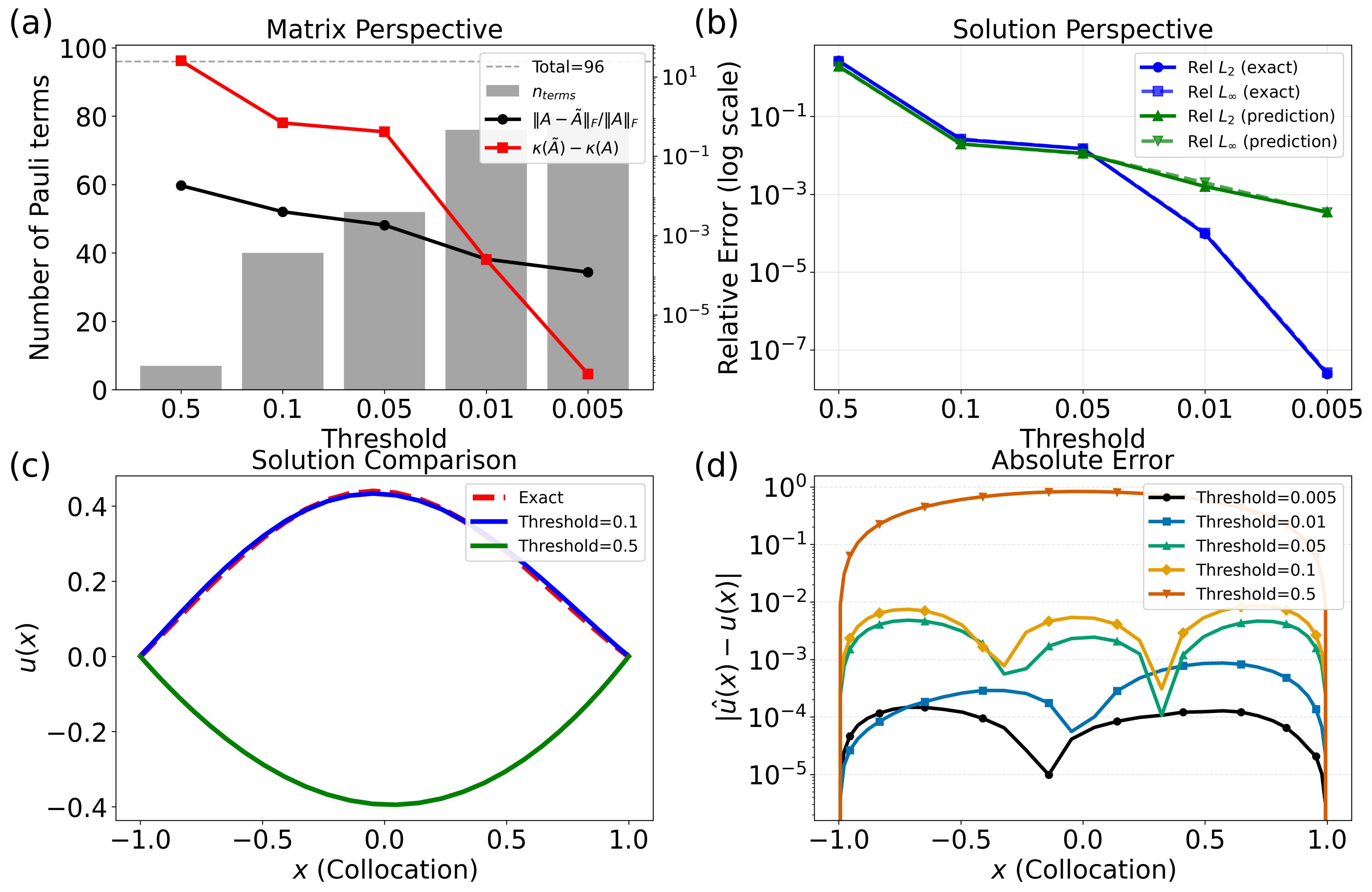}
\end{center}
\caption{Impact of the Pauli truncation threshold on the numerical properties of the truncated operator and the resulting solution accuracy for the one-dimensional Helmholtz equation with the homogeneous Dirichlet boundary condition and a wave number $k^2 = 4$.
Top: (a) Relative Frobenius-norm error between $A$ and $\tilde{A}$, condition number of $\tilde{A}$, and number of retained Pauli terms as functions of the truncation threshold.
(b) Solution error of the theoretically reconstructed solution and the solution predicted by our model.
Bottom: (c) Comparison of the solutions predicted by our model with the exact solution.
(d) Absolute error for different truncation thresholds.
}
\label{fig:Helmholtz_1D_Dirichlet_truncation_sim}
\end{figure}

We first analyze the effect of truncating the Pauli decomposition of $A$ by retaining terms in descending order of coefficient magnitude, where the truncation threshold is defined with respect to the coefficient size.
We consider the one-dimensional Helmholtz equation with Dirichlet boundary conditions and $k^2 = 4$ as a representative example, and summarize the results in Fig.~\ref{fig:Helmholtz_1D_Dirichlet_truncation_sim}. All learning experiments in this section are conducted using the same training setup as in Fig.~\ref{fig:Helmholtz_1D_Dirichlet}.

Figure~\ref{fig:Helmholtz_1D_Dirichlet_truncation_sim}(a) shows that lowering the truncation threshold reduces the relative approximation error between $A$ and the truncated operator $\tilde{A}$, measured in the Frobenius norm. At the same time, the condition number of $\tilde{A}$ approaches that of the full operator, while the number of retained Pauli terms increases. This indicates improved numerical conditioning of the truncated system with increased operator complexity.
Figure~\ref{fig:Helmholtz_1D_Dirichlet_truncation_sim}(b) compares the theoretically reconstructed solution obtained from $\tilde{\alpha}$ with the exact solution, together with the error of the solution predicted by our model using the truncated operator. For a threshold of $0.05$, the predicted solution closely follows the theoretical reconstruction, whereas smaller thresholds lead to noticeable deviations due to increased sensitivity of the learning problem. Nevertheless, at a threshold of $0.01$, the prediction error of our model remains on the order of $10^{-3}$, which is numerically acceptable for this problem.
Figure~\ref{fig:Helmholtz_1D_Dirichlet_truncation_sim}(c) and (d) further support these observations. Except for the most aggressive truncation case ($0.5$), the predicted solutions $(\leq 0.1)$ exhibit similar qualitative profiles to the exact solution. The corresponding spatial MAE highlights localized error amplification near the domain boundaries.

\begin{figure}[t]
\begin{center}
    \includegraphics[width=\linewidth]{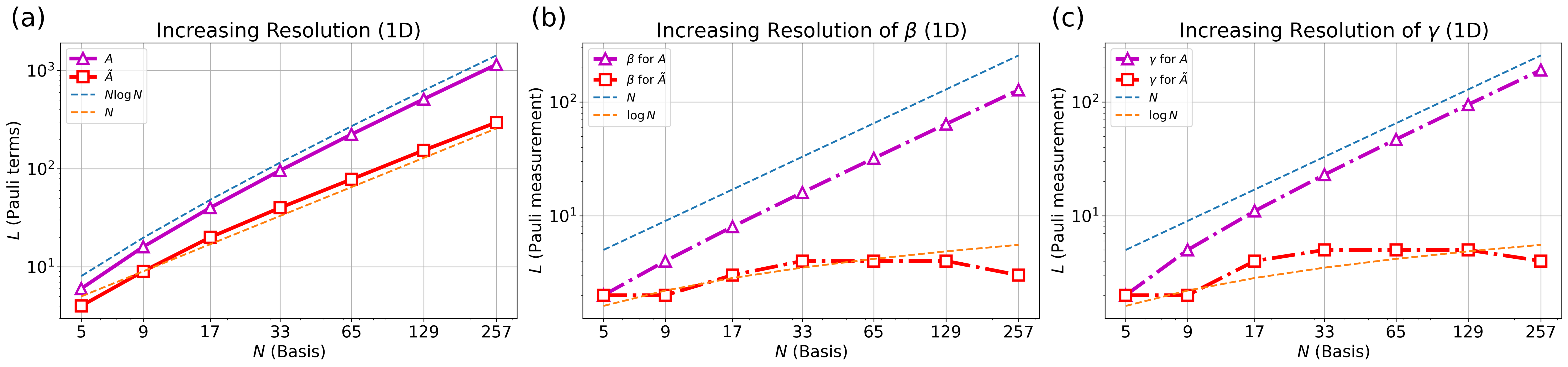}
\end{center}
\caption{Scaling behavior of the number of Pauli terms in the truncated operator $\tilde{A}$ and the corresponding measurement cost for loss evaluation in the one-dimensional Helmholtz equation with Dirichlet boundary conditions.
(a) Number of retained Pauli terms.
(b,c) Number of Pauli measurements required for evaluating $\beta$ and $\gamma$.
}
\label{fig:Helmholtz_1D_Dirichlet_truncation_number_of_Pauli}
\end{figure}

Based on the above analysis, we examine truncation thresholds that yield a theoretically induced solution error on the order of $10^{-3}$, which we regard as numerically acceptable for the present problem and comparable to the observed prediction accuracy of our model. Figure~\ref{fig:Helmholtz_1D_Dirichlet_truncation_number_of_Pauli} summarizes the resulting scaling behavior of the truncated operator. At this accuracy level, the number of retained Pauli terms in $\tilde{A}$ scales linearly with the system size $N^d$, reducing the original $\mathcal{O}(N^d \log N^d)$ complexity to $\mathcal{O}(N^d)$.

Moreover, the number of Pauli measurements required for evaluating the loss components is reduced from $\mathcal{O}(N^d)$ to $\mathcal{O}(\log N^d)$, leading to an overall logarithmic measurement complexity.
These results indicate that, when guided by an accuracy-based truncation criterion, substantial computational savings can be achieved while maintaining a solution accuracy that is sufficient for practical learning performance.

\begin{table}[t]
    \centering
    \caption{Performance of the NVQLS framework on 1D and 2D PDEs, including Reaction-diffusion (RD), Helmholtz, Convection-diffusion (CD), and Wave equation. Here, $N$ denotes the number of basis functions per dimension, and $n$ represents the required number of qubits.}
    \label{tab:steady_state_pdes}
    \begin{tabular}{llccccc}
        \toprule
        \multirow{2}{*}{\textbf{PDE}} & \multirow{2}{*}{\textbf{BC}} & 
        \multirow{2}{*}{ $N$($n$) } & \multicolumn{2}{c}{\textbf{Rel. $L_2$ Error}} & \multicolumn{2}{c}{\textbf{Rel. $L_\infty$ Error}} \\
        \cmidrule(lr){4-5} \cmidrule(lr){6-7}
        & & & Test Mean & Test SD & Test Mean & Test SD \\\midrule
        1D RD           & Dirichlet  & 32 (5) & $1.4 \times 10^{-3}$ & $1.7 \times 10^{-2}$ & $1.5 \times 10^{-3}$ & $1.7 \times 10^{-2}$ \\
        1D Helmholtz    & Neumann    & 32 (5) & $1.4 \times 10^{-2}$ & $5.7 \times 10^{-2}$ & $1.4 \times 10^{-2}$ & $5.7 \times 10^{-2}$ \\
        1D CD           & Dirichlet  & 32 (5) & $4.9 \times 10^{-3}$ & $3.5 \times 10^{-2}$ & $4.9 \times 10^{-3}$ & $3.1 \times 10^{-2}$ \\
        1D Wave         & Dirichlet  & 16 (4) & $1.6 \times 10^{-2}$ & $5.6 \times 10^{-2}$ & $1.1 \times 10^{-2}$ & $4.0 \times 10^{-2}$ \\
        \midrule
        2D RD           & Dirichlet  & 8 (6)  & $5.1 \times 10^{-3}$ & $5.6 \times 10^{-3}$ & $1.5 \times 10^{-2}$ & $8.2 \times 10^{-3}$ \\
        2D Helmholtz    & Neumann    & 8 (6)  & $3.0 \times 10^{-2}$ & $4.1 \times 10^{-2}$ & $3.2 \times 10^{-2}$ & $4.1 \times 10^{-2}$ \\
        2D CD           & Dirichlet  & 8 (6)  & $7.3 \times 10^{-3}$ & $5.3 \times 10^{-3}$ & $1.6 \times 10^{-2}$ & $8.9 \times 10^{-3}$ \\
        \midrule
        Joint Helmholtz & Dirichlet  & 8 (6)  & $3.4 \times 10^{-3}$ & $4.3 \times 10^{-3}$ & $3.7 \times 10^{-3}$ & $4.8 \times 10^{-3}$ \\
        \bottomrule
    \end{tabular}
    \label{tab:error_table}
\end{table}

\begin{table}[t]
    \centering
    \caption{Total execution time for the NVQLS framework across different PDE settings. Each experiment was conducted on a single NVIDIA RTX 3090 GPU.}
    \label{tab:execution_time}
    \begin{tabular}{lccc}
        \toprule
        \textbf{PDE} & $N$($n$) & \textbf{Epochs} & \textbf{Total Time (h)} \\
        \midrule
        1D RD           & 32 (5) & 30000  & 2.0 \\
        1D Helmholtz    & 32 (5) & 30000  & 1.9 \\
        1D CD           & 32 (5) & 30000  & 1.0 \\
        1D Wave         & 16 (4) & 50000  & 2.4 \\
        \midrule
        2D RD           & 8 (6)  & 200000 & 202.2 \\
        2D Helmholtz    & 8 (6)  & 200000 & 208.1 \\
        2D CD           & 8 (6)  & 200000 & 101.1 \\
        \midrule
        Joint Helmholtz & 8 (6)  & 100000 & 24.2 \\
        \bottomrule
    \end{tabular}
\end{table}

\section{Detailed Numerical Experiments}
\label{sec:experiment_details}

In this section, we provide the exact forms of the PDEs, their corresponding boundary conditions, and the resulting spectral matrices used in our experiments.

\subsection{1D Equations}

\textbf{1D Reaction--diffusion Equation.} We start with the one-dimensional reaction--diffusion equation with a diffusion coefficient $\epsilon$ under the homogeneous Dirichlet boundary condition:
\begin{equation}
\label{eq:RD1D_Dirichlet}
\begin{aligned}
    - \epsilon u_{xx}(x)+ u(x) &= f(x), 
    \quad x\in \Omega:=(-1, 1) \\
    u(x) &=0, \quad x \in \partial \, \Omega.
\end{aligned}
\end{equation}
Here, the Dirichlet conditions are realized via the basis functions
\begin{equation}
    \phi_k(x) = L_k(x) - L_{k+2}(x).
\label{eq:basis_Dirichlet}
\end{equation}

\textbf{1D Helmholtz Equation.}
The Helmholtz equation representing a wave propagation problem with a wave number $k$ is given by:
\begin{equation}
\label{eq:Helmholtz_1d_Neumann}
\begin{aligned}
    u_{xx}(x)+k^2 u(x)&=f(x)
    ,\quad
    x\in\Omega:=(-1,1)\\
    u_x(x) &=0, \quad
    x \in \partial \, \Omega.
\end{aligned}
\end{equation}
Here, the homogeneous Neumann boundary condition is strongly enforced by constructing the basis functions as
\begin{equation}
    \phi_k(x) = L_k(x) - \frac{k(k+1)}{(k+2)(k+3)} L_{k+2}(x).
\label{eq:basis_Neumann}
\end{equation}

\textbf{1D Convection--diffusion Equation.} We now focus on the convection--diffusion equation, characterized by a convection velocity $\nu$ and a small diffusion coefficient $\epsilon$:
\begin{equation}
\begin{aligned}
    - \epsilon u_{xx}(x) +\nu u_x(x) &= f(x),
    \quad x\in \Omega:=(-1, 1) \\
    u(x)&=0
    ,\quad
    x\in \partial \, \Omega.
\end{aligned}
\end{equation}

Based on the chosen basis functions, the corresponding spectral matrices for each equation are given by:
\begin{equation}
\begin{aligned}
    A_\text{RD 1D} 
    &= 
    - \epsilon S+M, \\
    A_\text{Helm 1D} 
    &= S+ k^2 M, \\
    A_\text{CD 1D} 
    &= 
    - \epsilon S+\nu R,
\end{aligned}
\end{equation}
where $S$, $M$, and $R$ denote the stiffness, mass, and convection matrices, respectively.

\begin{figure}[t]
\begin{center}
    \includegraphics[width=\linewidth]{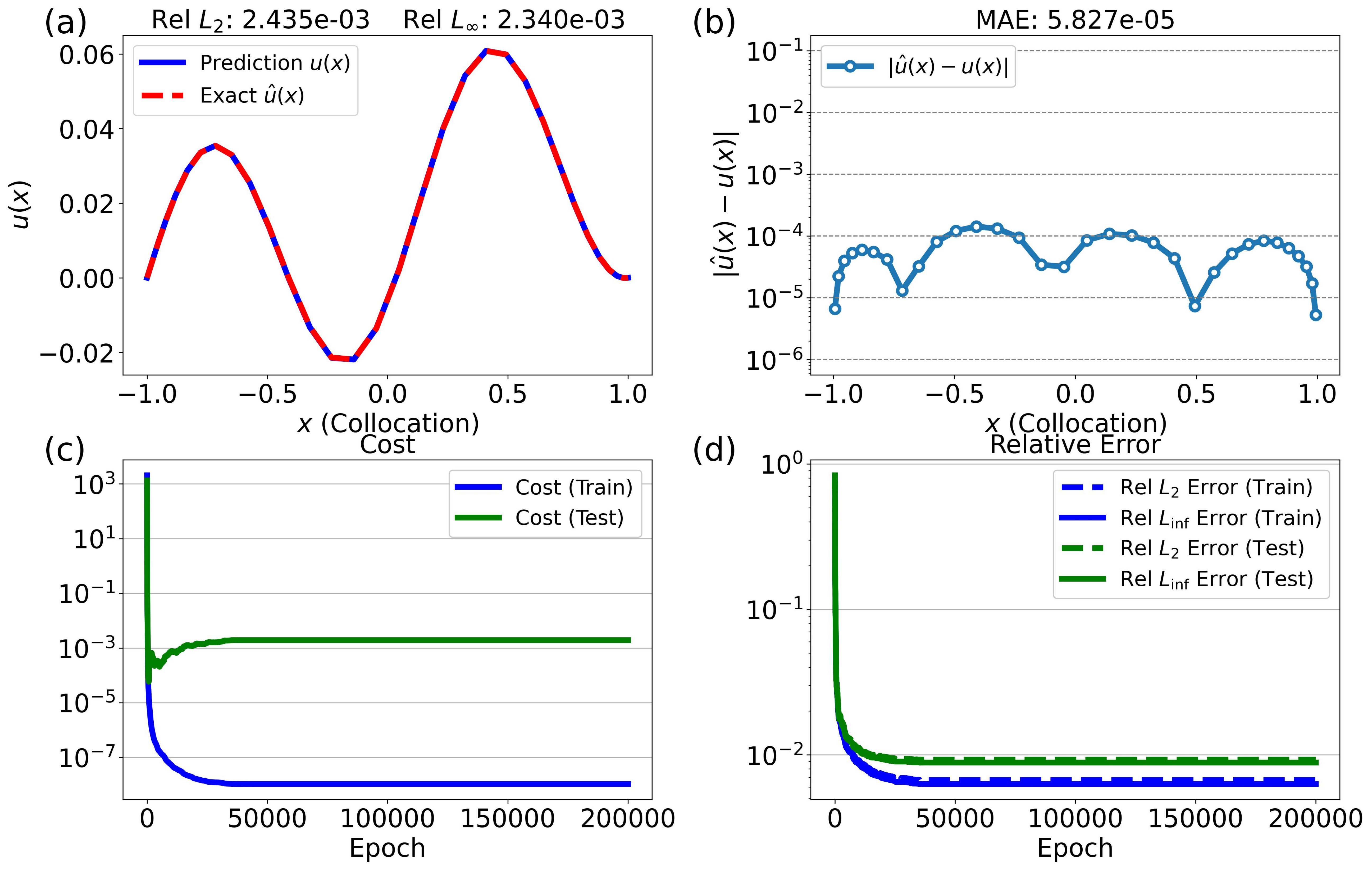}
\end{center}
\caption{
    Numerical results for the one-dimensional Helmholtz equation with the homogeneous Dirichlet boundary condition and a wave number $k^2=29.4$.
    Top:
    (a) example of the predicted solution $\hat{u}$ compared to the exact solution $u$,
    (b) absolute error between $\hat{u}$ and $u$.
    Bottom:
    (c) batch-wise training and test losses,
    (d) batch-wise relative $L^2$ and $L^\infty$ errors over epochs.
}
\label{fig:Helmholtz_1D_Dirichlet}
\end{figure}

\textbf{Helmholtz Equation (Dirichlet Boundary).}
To demonstrate the ability of the proposed method to handle various boundary conditions, we focus on the Helmholtz equation with another boundary condition---Dirichlet BC. The exact form of the equation is given by:
\begin{equation}
\label{eq:Helmholtz_1d_Dirichlet}
\begin{aligned}
    u_{xx}(x)+k^2 u(x)&=f(x)
    ,\quad
    x\in\Omega:=(-1,1)\\
    u(x) &=0, \quad
    x \in \partial \, \Omega.
\end{aligned}
\end{equation}
The corresponding matrix is $A=S+k^2M$, which is the same matrix as the Neumann BC. Given a wave number $k^2=29.4$, Fig.~\ref{fig:Helmholtz_1D_Dirichlet} (a) compares the exact solution with the model’s prediction, showing a relative error below $0.3\%$, which confirms their strong consistency. The corresponding absolute error is depicted in Fig.~\ref{fig:Helmholtz_1D_Dirichlet} (b), yielding an MAE of $5.827\times 10^{-5}$. 
Figure~\ref{fig:Helmholtz_1D_Dirichlet} (c)--(d) track the training and test dynamics, including both the loss curves and the relative errors. A steady decrease in relative error accompanies the reduction of training cost. Even though the test loss begins to rise early, the continued decrease in the test relative error indicates that further training remains beneficial for achieving better solution accuracy.

\begin{figure}[t]
\begin{center}
    \includegraphics[width=\linewidth]{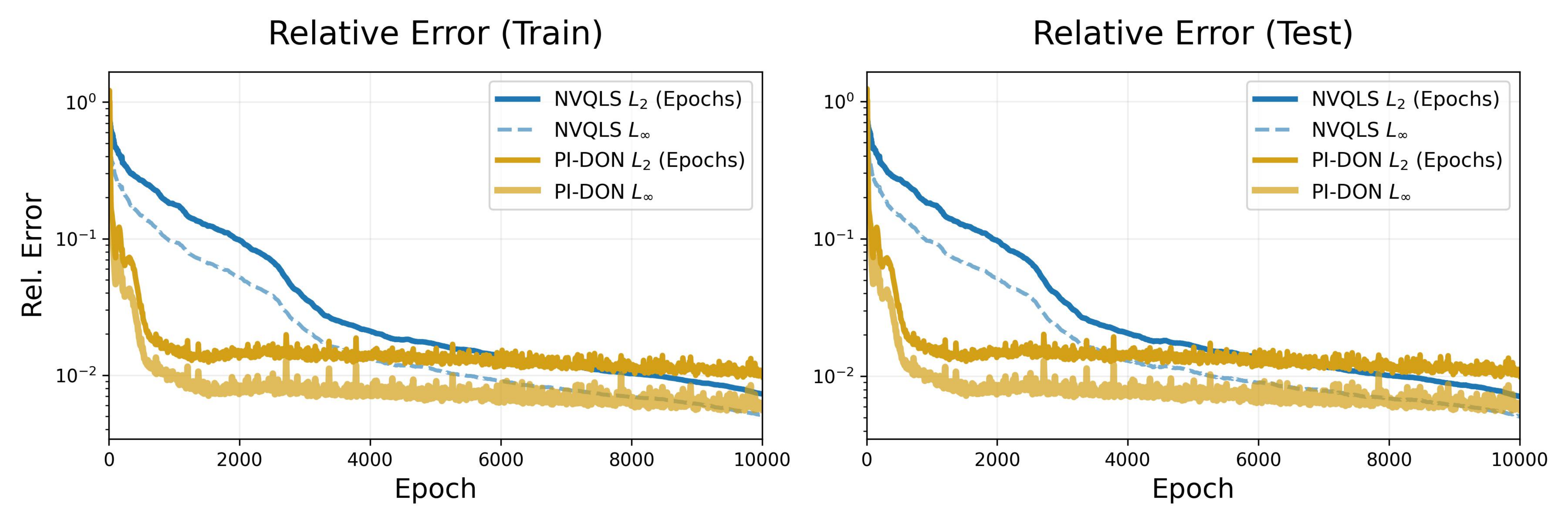}
\end{center}
    \caption{Visualization of the training and testing error curves for the proposed NVQLS and the PI-DON baseline on the 1D wave equation.}
\label{fig:comparison_wave}
\end{figure}

\textbf{1D Wave Equation.} We trained NVQLS on the 1D wave equation that is time-dependent and hyperbolic, defined as 
\begin{equation}
    u_{tt} - u_{xx} = f(x,t), \quad (x,t) \in [0, 1] \times [0, T],
\end{equation}
with the homogeneous Dirichlet boundary conditions and zero initial conditions (i.e., $u(x, 0) = u_t(x, 0) = 0$). Here, we utilized the external forcing functions, which are parametrized by
\begin{equation}
\label{eq:wave_forcing}
    f(x,t) = \left[ 1 + \frac{\pi^2(1+\omega)^2 t^2}{2}  \right] \sin(\pi(1+\omega)x) + \left[ 1 + \frac{\pi^2(1-\omega)^2 t^2}{2}  \right] \sin(\pi(1-\omega)x),
\end{equation}
where $\omega$ is sampled from a uniform distribution $U[1, 2)$. This setup evaluates the model's ability to learn a family of dynamic wave solutions with varying frequencies. The corresponding spectral matrix for the wave equation is given by:
\begin{equation}
    A_\text{Wave 1D} = M\otimes S- S \otimes M.
\end{equation}

\subsection{2D Equations}

The remaining sections of the numerical experiments focus on results for various two-dimensional problems. For two-dimensional domains, the solution is expressed as a linear combination of basis functions constructed via the tensor product of the corresponding one-dimensional bases in $x$ and $y$:
\begin{equation}
      u(x,y) = \sum_{k,j=0}^{N-1}
\alpha_{kj}
\phi_k(x)\phi_j(y).
\end{equation}
    Similar to the one-dimensional cases, NVQLS is trained in an unsupervised manner to predict the corresponding solutions from the forcing function inputs in the two-dimensional setting.
    
    \textbf{2D Reaction--diffusion Equation.} We first extend our study to the two-dimensional reaction--diffusion equation with a diffusion coefficient $\epsilon$, subject to homogeneous Dirichlet boundary conditions:
\begin{equation}
    \begin{aligned}
    -\epsilon \Delta u(x,y)
    + u(x,y)
    &= f(x,y)
    ,\quad
    (x,y) \in \Omega
    \\
    u(x,y)&=0
    , \quad
    (x,y) \in \partial \, \Omega.
\end{aligned}
\end{equation}

\textbf{2D Helmholtz Equation.} Next, we present the numerical setup for the two-dimensional Helmholtz equation. The specific form of the equation is characterized by the wave number $k$, given by:
\begin{equation}
\begin{aligned}
    \Delta u(x,y)
    + k^2 u(x,y)
    &= f(x,y)
    ,\quad
    (x,y) \in \Omega
    \\
    u(x,y) &=0
    ,\quad
    (x,y) \in \partial \, \Omega.
\end{aligned}
\end{equation}

\textbf{2D Convection--diffusion Equation.} Finally, we consider the convection--diffusion equation with homogeneous Dirichlet boundary conditions in two dimensions. The exact form is defined with a small diffusion coefficient $\epsilon$ and a convection velocity vector $\nu = (\nu_1, \nu_2)^T$:
\begin{equation}
\begin{aligned}
    -\epsilon \Delta u(x,y)
    + \nu \cdot \nabla u(x,y)
    &= f(x,y)
    ,\quad (x,y) \in \Omega \\
    u(x,y) &= 0
    ,\quad (x,y) \in \partial \, \Omega.
\end{aligned}
\end{equation}

Based on the weak formulation and the tensor product structure, the corresponding spectral matrices for the 2D PDEs are formulated as:
\begin{equation}
\begin{aligned}
    A_\text{RD 2D} &= 
    -\epsilon \left( 
    S\otimes M
    + M\otimes S \right)
    + M \otimes M \\
    A_\text{Helm 2D} &= 
    S\otimes M
    + M\otimes S
    + k^2 M \otimes M\\
    A_\text{CD 2D} &= 
    -\epsilon \left( 
    S\otimes M
    + M\otimes S \right)
    +\nu_1 \, R \otimes M
    +\nu_2 \, M \otimes R^T.
\end{aligned}
\end{equation}

\begin{figure}[t]
\begin{center}
    \includegraphics[width=\linewidth]{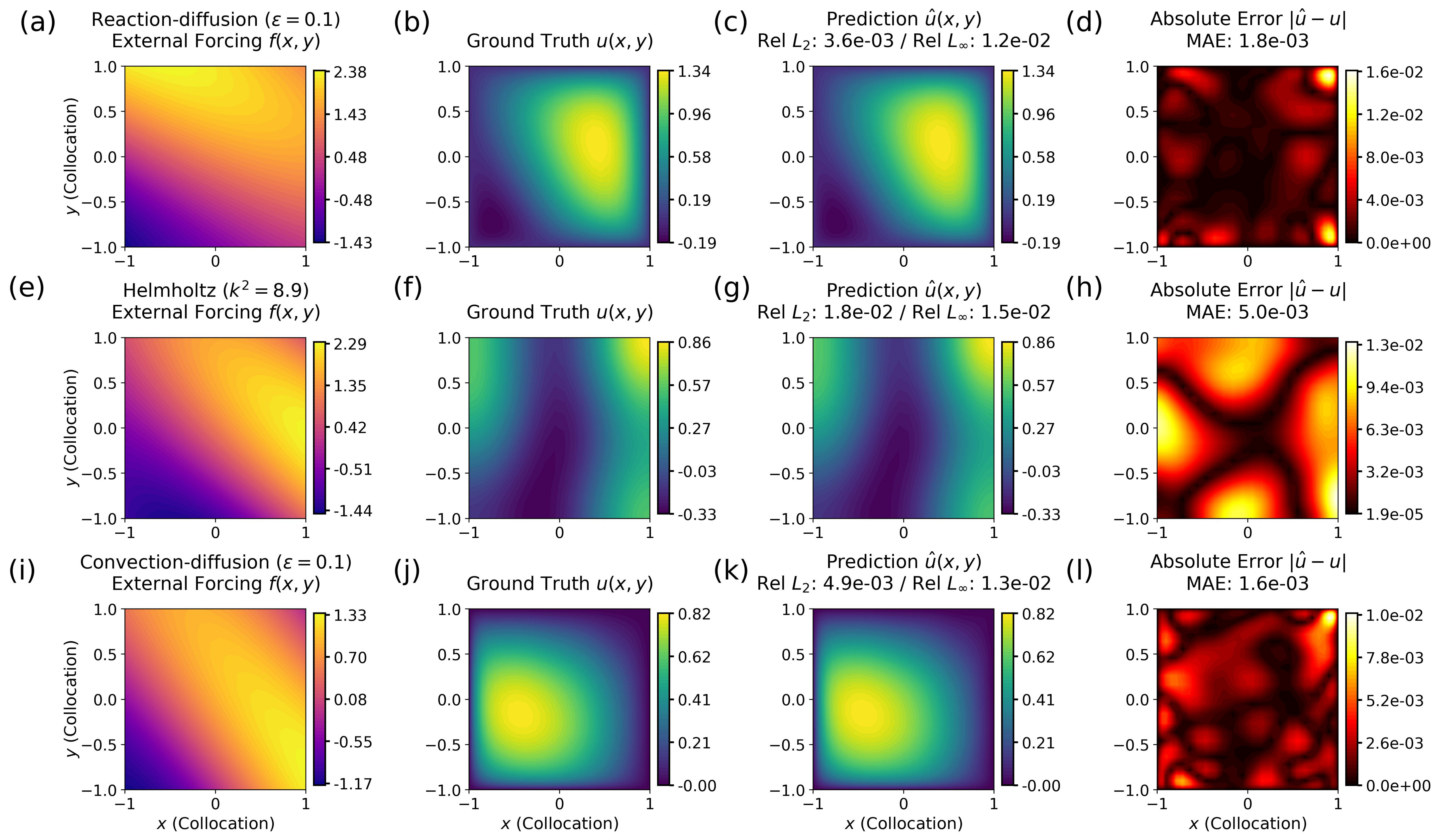}
\end{center}
\caption{
    Numerical results for two-dimensional PDEs. 
    Columns from left to right display the external forcing $f$, the exact ground truth solution $u$, the NVQLS prediction $\hat{u}$, and the pointwise absolute error $|\hat{u}-u|$, respectively. 
    Rows from top to bottom correspond to the evaluated physical systems: (a)-(d) the reaction--diffusion equation with $\epsilon=0.1$ (Dirichlet BC), (e)-(h) the Helmholtz equation with $k^2=8.9$ (Neumann BC), and (i)-(l) the convection--diffusion equation with $\epsilon=0.1$ (Dirichlet BC).
    }
\label{fig:2d_pdes_example}
\end{figure}

Figure~\ref{fig:2d_pdes_example} illustrates the representative numerical results for these two-dimensional benchmarks, comparing the exact solutions against the NVQLS predictions. Specifically, panels (a), (b), (c), and (d) present the external forcing, the exact solution, the predicted solution, and the pointwise absolute error, respectively, for the reaction--diffusion equation with $\epsilon=0.1$. Similarly, panels (e), (f), (g), and (h) display the corresponding results for the Helmholtz equation with $k^2=8.9$. Finally, panels (i), (j), (k), and (l) show the results for the convection--diffusion equation with a diffusion parameter $\epsilon=0.1$. Across all three physical systems, the panels reveal a strong agreement between the predictions and the exact solutions. The relative $L^2$ errors for all benchmarks are consistently maintained at the order of $10^{-2}$ or $10^{-3}$. Notably, the framework achieves high fidelity with MAE values as low as $1.6 \times 10^{-3}$ for the convection--diffusion equation, demonstrating the robustness of our framework for 2D spectral operator learning.

\subsection{Spectral Operator Learning with Joint Coefficient-Forcing Inputs}
\label{sec:parametric_1d}

\begin{figure}[t]
\begin{center}
    \includegraphics[width=\linewidth]{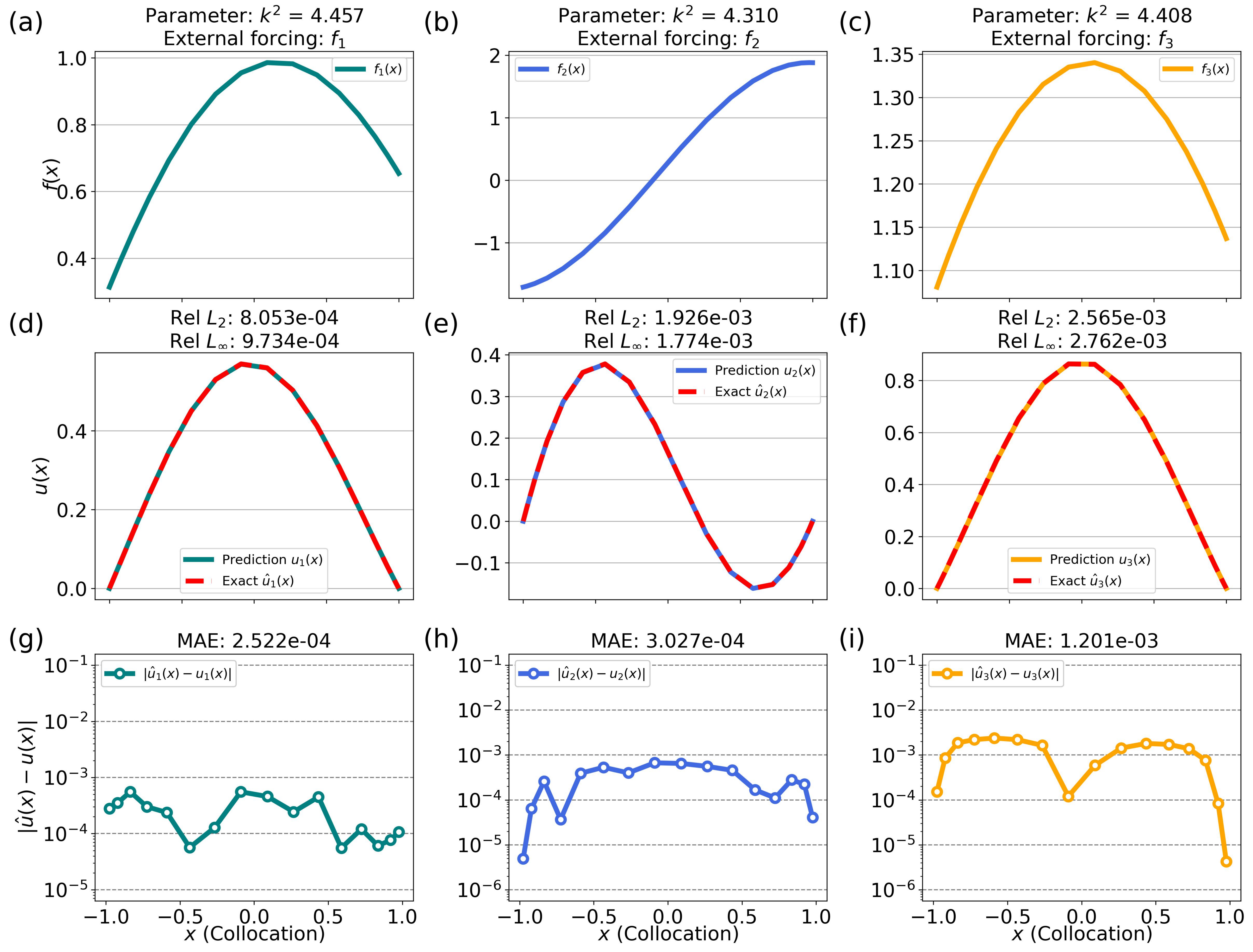}
\end{center}
    \caption{
    Numerical examples of operator learning with joint parameter and forcing inputs for the Helmholtz equation under Dirichlet boundary conditions.
    Top:
    input pairs for the angle network:
    (a) case 1: a forcing function $f_1$ and a wave number $k^2=4.457$,
    (b) case 2: $f_2$ and $k^2=4.310$,
    (c) case 3: $f_3$ and $k^2=4.408$.
    Middle:
    exact solutions $u_i$ and predicted solutions $\hat{u}_i$ for each case:
    (d) case 1,  
    (e) case 2,  
    (f) case 3.
    Bottom:
    absolute error plots $|\hat{u}_i - u_i|$ on collocation points excluding the boundary points
    (g) case 1,
    (h) case 2,
    (i) case 3.
    }
\label{fig:general_Helmholtz_1d}
\end{figure}

This section introduces operator learning with joint parameter and forcing inputs, a framework in which the learned operator maps both a forcing function and a PDE parameter to the corresponding PDE solution. We demonstrate this approach using the Helmholtz equation with a homogeneous Dirichlet boundary condition given in Eq.~\eqref{eq:Helmholtz_1d_Dirichlet}.

To implement operator learning with joint inputs, the input instances are tuples of a forcing function $f^{(i)}$ and a wave number $k^{(i)}$, represented as $\{ (k^{(i)}, f^{(i)}) \}_{i=1}^D$ and mapped to angle parameters $\theta^{(i)} = g(k^{(i)}, f^{(i)}; w)$ via the angle network $g$. Given an input instance $(k,f)$, the spectral method matrix for the Helmholtz equation, $A(k)$, is determined by the parameter $k$ and expressed as a linear combination of the fixed stiffness and mass matrices, $S$ and $M$. We next perform Pauli decompositions of these fixed matrices where the Pauli terms $P_{l_1}$ and $P_{l_2}$ can represent the same operator:
\begin{equation}
\label{eq:joint_Helmholtz_1d}
    A = S + k^2 M
    =
    \sum_{l_1} c_{l_1} P_{l_1}
    +  k^2
    \sum_{l_2} c_{l_2} P_{l_2}.
\end{equation}

Since the fixed matrices $S$ and $M$ are independent of $k$, their Pauli decompositions need only be computed once and can be reused for any value of $k$. Denoting $\ket{\hat{\alpha}} = \ket{\hat{\alpha}(k, f;w)}$, we finally express the numerator of the proposed cost function in Eq.~\eqref{eq:loss_NVQLS} as:
\begin{equation}
\label{eq:parametric_beta_1d}
    \sum_{l_1}
\mathrm{Re}\left( c_{l_1} \bra{F}
     P_{l_1}
    \ket{{\hat{\alpha}}} \right)
    + k^2
    \sum_{l_2}
\mathrm{Re}\left( c_{l_2} \bra{F}
    P_{l_2}
    \ket{{\hat{\alpha}}} \right).
\end{equation}

Similarly, the term inside the square root in the denominator of the cost function is given by the following expression, where Pauli terms $P_{l_3}$, $P_{l_4}$, $P_{l_5}$, and $P_{l_6}$ represent the matrices $S^\dagger S$, $S^\dagger M$, $M^\dagger S$, and $M^\dagger M$, respectively, with their corresponding coefficients:
\begin{equation}
\label{eq:parametric_gamma_1d}
    \sum_{l_3}
    c_{l_3}
    \bra{\hat{\alpha}}
    P_{l_3}
    \ket{\hat{\alpha}}
    + k^2
    \sum_{l_4}
    c_{l_4}
    \bra{\hat{\alpha}}
    P_{l_4}
    \ket{\hat{\alpha}}
    + k^2
    \sum_{l_5}
    c_{l_5}
    \bra{\hat{\alpha}}
    P_{l_5}
    \ket{\hat{\alpha}}
    + k^4
    \sum_{l_6}
    c_{l_6}
    \bra{\hat{\alpha}}
    P_{l_6}
    \ket{\hat{\alpha}}.
\end{equation}

Figure~\ref{fig:general_Helmholtz_1d} presents three representative examples of operator learning with joint parameter and forcing inputs for the Helmholtz equation, using Dirichlet boundary conditions. The top row, shown in Fig.~\ref{fig:general_Helmholtz_1d}(a)--(c), presents three input instances, $(k_1, f_1)$, $(k_2, f_2)$, and $(k_3, f_3)$, where the wave numbers $k_i$ are sampled from the uniform distribution $U[4, 5)$ and the forcing functions are linear combinations of trigonometric functions. The middle and bottom rows (Fig.~\ref{fig:general_Helmholtz_1d}(d)--(f) and (g)--(i), respectively) present NVQLS predictions versus exact solutions and absolute differences along with MAE values. The predictions exhibit small relative errors (below $0.3\%$) and MAE values (below $2\times10^{-3}$), demonstrating the model’s ability to approximate the solution operator across varying parameters.


\end{document}